%% file: main.tex
\documentclass[conference]{IEEEtran}
\usepackage{cite}
\usepackage{amsmath,amssymb,amsfonts}
\usepackage{algorithmic}
\usepackage{graphicx}
\usepackage{textcomp}
\usepackage{xcolor}
\usepackage[hyphens]{url}

\def\BibTeX{{\rm B\kern-.05em{\sc i\kern-.025em b}\kern-.08em
    T\kern-.1667em\lower.7ex\hbox{E}\kern-.125emX}}

\usepackage{mathptmx} 
\usepackage{amsmath,amssymb,amsfonts}
\usepackage{graphicx}
\usepackage{textcomp}
\usepackage{xcolor}
\usepackage{color,soul}
\usepackage{booktabs}
\usepackage{tabularx}
\usepackage{balance}
\usepackage{amsbsy}
\usepackage{pifont}
\usepackage{listings}
\usepackage{xspace}
\usepackage[normalem]{ulem}

\usepackage{cleveref}
\crefformat{figure}{Fig.~#2#1#3}
\crefformat{chapter}{\S#2#1#3} 
\crefformat{section}{\S#2#1#3} 
\crefformat{subsection}{\S#2#1#3}
\crefformat{subsubsection}{\S#2#1#3}

\definecolor{darkred}{rgb}{0.8, 0.0, 0.0}
\definecolor{darkgreen}{rgb}{0.0, 0.5, 0.0}
\definecolor{mustard}{rgb}{1.0, 0.77, 0.05}
\newcommand{\cmark}{\ding{51}}%
\newcommand{\xmark}{\ding{55}}%

\newcommand{\proj}{DCRA\xspace}
\newcommand{\device}{DCRA\xspace}

\newif\ifarxiv
\arxivtrue 

\author{
Marcelo Orenes-Vera, Esin Tureci, Margaret Martonosi, David Wentzlaff\\
Princeton University, Princeton, New Jersey, USA \\
\{movera, esin.tureci, mrm, wentzlaf\} @princeton.edu
}
\title{DCRA: A Distributed Chiplet-based Reconfigurable Architecture for Irregular Applications}
\newcommand{\repository}{repository\cite{dcra_repo_names}\xspace}

\begin{document}
\date{}
\maketitle
\pagestyle{plain}

\newcommand{\subf}[2]{%
  {\small\begin{tabular}[t]{@{}c@{}}{\tiny}
  #1\\#2
  \end{tabular}}%
}

\input{tex/dcra.tex}

\section*{Acknowledgments}
\noindent
This material is based on research sponsored by the Air Force Research Laboratory (AFRL), Defense Advanced Research Projects Agency (DARPA) under agreement FA8650-18-2-7862, and National Science Foundation (NSF)
award No. 1763838.~\footnote{The U.S. Government is authorized to reproduce and distribute reprints for Governmental purposes notwithstanding any copyright notation thereon. The views and conclusions contained herein are those of the authors and should not be interpreted as necessarily representing the official policies or endorsements, either expressed or implied, of NSF, AFRL and DARPA or the U.S. Government.}

\bibliographystyle{plain}
\bibliography{refs}

\end{document}

%% file: tex/dcra.tex
\begin{abstract}

In recent years, the growing demand to process large graphs and sparse datasets has led to increased research efforts to develop hardware- and software-based architectural solutions to accelerate them.
While some of these approaches achieve scalable parallelization with up to thousands of cores, adaptation of these proposals by the industry remained slow.
To help solve this dissonance, we identified a set of questions and considerations that current research has not considered deeply.

Starting from a tile-based architecture, we put forward a Distributed Chiplet-based Reconfigurable Architecture (DCRA) for irregular applications that carefully consider fabrication constraints that made prior work either hard or costly to implement or too rigid to be applied.
We identify and study pre-silicon, package-time and compile-time configurations that help optimize DCRA for different deployments and target metrics.
To enable that, we propose a practical path for manufacturing chip packages by composing variable numbers of DCRA and memory dies, with a software-configurable Torus network to connect them.
We evaluate six applications and four datasets, with several configurations and memory technologies, to provide a detailed analysis of the performance, power, and cost of DCRA as a compute node for scale-out sparse data processing.
Finally, we present our findings and discuss how DCRA's framework for design exploration can help guide architects to build scalable and cost-efficient systems for irregular applications.

\end{abstract}

\section{Introduction}

Irregular applications are those traversing data structures such as trees, graphs, and sparse matrices, whose data-dependent access patterns lead to unpredictable memory accesses.
They include applications doing graph analytics, sparse tensor algebra, particle simulations, and database operations.
The increasing relevance of these applications has led the computer architecture community to develop many promising software and hardware techniques to accelerate them~\cite{tesseract,fifer,polygraph,dalorex,jeffrey_hive,pipette,maple,graphcore,spade}.
However, there is still an unmet demand for systems that can accelerate these applications at scale~\cite{iarpa_agile,agile_hpcwire}.

Motivated by the idea of building such systems, we faced the following research questions:
(a) what is the design space of architectures for irregular applications;
(b) how do different target metrics (e.g., time-to-solution, energy, cost) affect design choices; and
(c) how different applications and datasets are affected by these decisions.

\begin{figure}[t]
\centering  
\vspace{-3mm}
\includegraphics[width=\columnwidth]{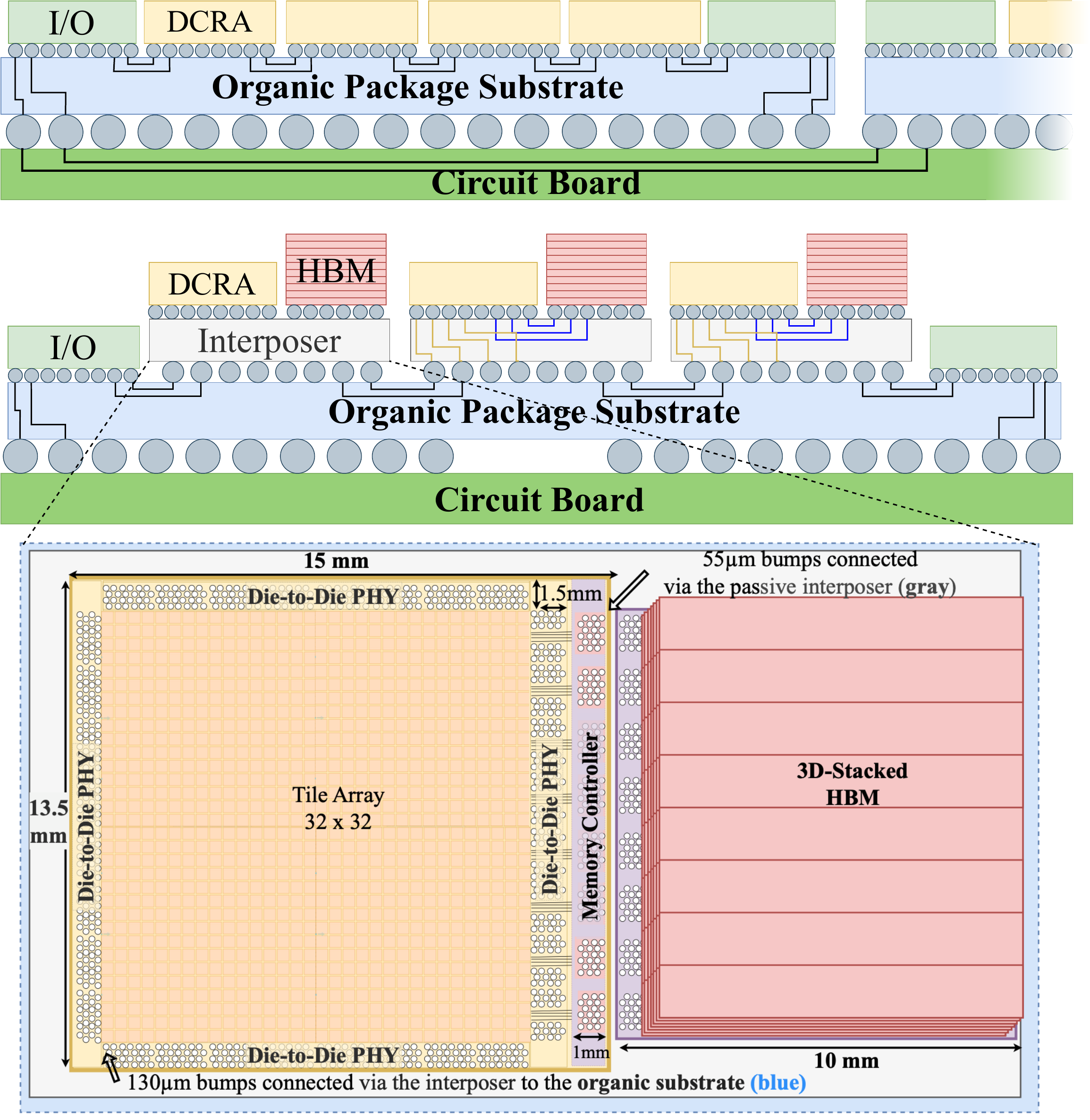}
\vspace{-4mm}
\caption{
Two possible integrations of the same 32x32-tile \device die (dimensions depicted with 256KB SRAM/tile).
Top: two packages on a board, each featuring only \device dies, optimized for time-to-solution as it maximizes parallelization (lower data footprint per die).
Bottom: a single package with \device dies and stacked DRAM, optimized for performance-per-dollar. 
}
\vspace{-2mm}
\label{fig:cake}
\end{figure}

Prior work Dalorex managed to scale billion-edge graphs across thousands of cores by having the entire graph stored on-chip.
It followed a similar approach to the Cerebras Wafer-Scale Engine~\cite{cerebras} of having a huge amount of SRAM on-chip, distributed evenly across a tiled grid of processing elements (PEs).
While this approach achieves the largest memory bandwidth per PE---a precious resource for data-intensive applications---it sets the ratio of compute and memory per PE at silicon time, which then constrains the minimal level of parallelization at which to run a given dataset.
Memory bandwidth is only part of the solution to scaling out irregular applications, as the network also plays a key role when scaling to a large number of PEs.
Building large networks on a 2D silicon limits the options to high-diameter 2D topologies, which only scale bisection bandwidth---also a precious resource---with the square root of the number of PEs.
Thus, massively scaling irregular applications would eventually require having chips distributed on a more beefy network in the 3D space~\cite{dragonfly,fattree,polarfly} and using problem partitioning to create locality within each node~\cite{metis,parmetis,scotch}.
But how big each node should be and the design tradeoffs to consider when architecting it, are open questions that we aim to answer in this paper.

Despite the large corpus of work on irregular memory access patterns and graph processing, there is still a lack of studies on the limits of parallelization within each node and the tradeoffs between performance, power, area (PPA) and cost involved in building it.

\textbf{Our work}:
To answer these research questions, we revisit recent works in manycore architectures and graph processing through the lens of modern technology trends to:
(1) understand the manufacturing process and constraints of the architectures proposed by prior work;
(2) study how the balance of hardware resources dedicated to compute, memory, and network affects the energy, performance, and cost of running irregular applications;
(3) propose \proj, a \textul{d}istributed \textul{c}hiplet-based \textul{r}econfigurable \textul{a}rchitecture that allows taking some of these design decisions post-silicon---at packaging time;
(4) analyze several pre-silicon options and packaging- and compile-time configurations of \proj in terms of performance, energy and cost efficiency for six irregular applications.

Designing computing nodes composed of tens of thousands of cores requires taking cost constraints into account.
We investigated the fabrication cost and resource utilization of monolithic architectures like Dalorex and found that \textit{\textbf{it is hard to strike a balance for the amount of SRAM per PE}} tile.
A small SRAM per tile results in higher peak compute throughput for the same area, but it forces the spread of large datasets across more PEs, putting more pressure on the network and resulting in low utilization, for communication-intensive applications.
In contrast, with a large SRAM per tile, the dataset can be accommodated across fewer tiles and get closer to the ideal compute utilization, but the performance per dollar will be lower when we want to achieve maximum parallelization.
A similar problem applies to processing-in-memory (PIM) architectures like Tesseract~\cite{tesseract} and GraphQ~\cite{2018graphp}; the ratio of PEs per DRAM bank is fixed during fabrication, leading to a similar tradeoff at a different scale.

Because the cost of fabricating separate silicon dies for products with different performance targets could thwart performance-per-dollar goals, we propose \proj, a composable chiplet architecture where crucial design decisions like the ratio of memory to compute resources, total on-chip memory capacity or off-chip bandwidth are taken more cost-efficiently, during chip-packaging.

\cref{fig:cake} depicts two packaging configurations where the same \device{} dies are integrated to create different products, each optimized for a target metric, i.e., time-to-solution (above) and performance-per-dollar (below).
DRAM devices can be interleaved between \device{} dies to increase the capacity of a tile's local memory since the DRAM storage is exclusively assigned to the tiles of the adjacent \device{} die.
The IO dies on the sides of the package determine the off-chip bandwidth. Multiple chips could be integrated into a board, which would define what we consider a \emph{compute node}.

The rest of the paper is organized as follows.
\cref{sec:background} reviews the state-of-the-art in manycore and graph architectures and motivates why a chiplet-based reconfigurable architecture can help meet different PPA targets cost-effectively for irregular applications.
\cref{sec:approach} describes the \proj architecture, including our design contributions towards supporting nodes of configurable sizes optimized for irregular applications, where memory and compute are interleaved on-chip, connected by a software-defined Torus network.
It also describes the design innovations that enabled some of these configurations.
\cref{sec:methodology} describes the chiplet model and the cost model that we built on top of the simulator used for Dalorex, and the applications and datasets used in our evaluation.
\cref{sec:results} presents our characterizations of the \proj pre-silicon options, as well as the post-silicon and software-time configurations, and shows the performance-per-dollar and energy-efficiency benefits of \proj over prior work.
Before concluding, \cref{sec:discussion} presents the insights we gained from these characterizations and provides guidance for future designs.

\section{Background and Motivation}\label{sec:background}

Memory accesses from graph and sparse linear algebra applications exhibit little spatial or temporal locality, resulting in poor cache behavior and intense traffic throughout the memory hierarchy~\cite{graphattack}.
Prior work aiming to accelerate these workloads mitigate memory latency via decoupling, prefetching and pipelining techniques~\cite{prodigy,maple,graphicionado,graphpulse,ozdal,chronos, pipette, fifer, polygraph, jeffrey_hive, jeffrey_swarm,cohort}.
Fifer~\cite{fifer} and Polygraph~\cite{polygraph} increase utilization further through spatiotemporal parallelization, while Hive~\cite{jeffrey_hive} provides ordered parallelization.
These works improve the execution of irregular applications via hardware-software co-designs, without exploring much of the hardware design space in terms of balancing compute and memory resources.
Tesseract~\cite{tesseract} and GraphQ~\cite{2019graphq} explored tight integrations of compute and memory resources by integrating PEs on the logic layer of a 3D stacked memory~\cite{hmc}, while Dalorex~\cite{dalorex} proposes having all the memory on SRAM, scattered evenly across the PEs, to increase even further the memory bandwidth.
In addition, Dalorex builds on prior software techniques and adds hardware support to offer a programming model where the program is split into tasks at pointer indirection so that all tasks execute on the PEs that have the data in their local memory.
Thanks to all of this prior work, as a community, we have a better understanding of how to map and orchestrate irregular applications to the hardware.
We argue that the next step toward realizing some of these hardware-software innovations at scale is \textit{\textbf{exploring the architecture design space while considering physical implementation constraints and cost}}.

For the rest of this section, we explain how chiplets can help with cost-effectiveness beyond improving silicon yield (\cref{sec:background_nre}) and review the state-of-the-art in manycore architectures (\cref{sec:background_manycore}) to motivate the need for a new reconfigurable architecture for irregular applications.

\setlength{\tabcolsep}{1.5pt}
\begin{table}[t]
\caption{Post-silicon reconfigurability of the network-on-chip (NoC), memory capacity and \# of processing elements (PEs) per chip, and whether the PEs can execute software instructions.}
\centering
\small
\begin{tabularx}{\columnwidth}{@{\hspace{-1pt}}l@{\hspace{1pt}}c@{\hspace{2pt}}c@{\hspace{2pt}}c@{\hspace{1pt}}c@{\hspace{1pt}} }
\toprule
 \textbf{Reconfigurable}                                   & Memory               &  \#Processing                       &  Configurable            & Software ISA \\
 \textbf{Aspects} /                                      & Size/BW            &  Elements                             &  NoC                 & Programmable \\
 Prior Work                                              & per chip            &  per chip                          &  Topology                & Processors \\
\midrule
Fifer \cite{esperanto}                                & {\color{mustard}\pmb{\cmark}} & {\color{darkred}\pmb{\xmark}} & {\color{darkred}\pmb{\xmark}} & {\color{darkred}\pmb{\xmark}} \\ 
Tesseract~\cite{tesseract}   & {\color{darkgreen}\pmb{\cmark}} & {\color{darkred}\pmb{\xmark}} & {\color{darkred}\pmb{\xmark}} & {\color{darkgreen}\pmb{\cmark}} \\
Dalorex \cite{dalorex}                                & {\color{darkred}\pmb{\xmark}} & {\color{darkred}\pmb{\xmark}} & {\color{darkred}\pmb{\xmark}} & {\color{darkgreen}\pmb{\cmark}} \\
PolyGraph \cite{polygraph}                            & {\color{mustard}\pmb{\cmark}} & {\color{darkred}\pmb{\xmark}} & {\color{darkred}\pmb{\xmark}} & {\color{darkred}\pmb{\xmark}} \\
Decades~\cite{piton+ariane,decades}                   & {\color{darkred}\pmb{\xmark}} & {\color{darkred}\pmb{\xmark}} & {\color{darkred}\pmb{\xmark}} & {\color{darkgreen}\pmb{\cmark}} \\
\midrule
ESP~\cite{giri_date20}                                & {\color{darkred}\pmb{\xmark}} & {\color{darkred}\pmb{\xmark}} & {\color{darkred}\pmb{\xmark}} & {\color{darkgreen}\pmb{\cmark}} \\
Manticore \cite{manticore}                            & {\color{darkgreen}\pmb{\cmark}} & {\color{darkred}\pmb{\xmark}} & {\color{darkred}\pmb{\xmark}} & {\color{darkgreen}\pmb{\cmark}} \\
Simba~\cite{simba}                                    & {\color{darkred}\pmb{\xmark}} & {\color{darkgreen}\pmb{\cmark}} & {\color{mustard}\pmb{\cmark}} & {\color{darkgreen}\pmb{\cmark}} \\
Intel's PIUMA \cite{piuma}                            & {\color{darkgreen}\pmb{\cmark}} & {\color{darkred}\pmb{\xmark}} & {\color{darkred}\pmb{\xmark}} & {\color{darkgreen}\pmb{\cmark}} \\
Graphcore \cite{graphcore}                            & {\color{darkred}\pmb{\xmark}} & {\color{darkred}\pmb{\xmark}} & {\color{mustard}\pmb{\cmark}} & {\color{mustard}\pmb{\cmark}} \\
Sambanova \cite{sambanova}                            & {\color{mustard}\pmb{\cmark}} & {\color{mustard}\pmb{\cmark}} & {\color{darkgreen}\pmb{\cmark}} & {\color{darkred}\pmb{\xmark}} \\
Cerebras \cite{cerebras}                              & {\color{darkred}\pmb{\xmark}} & {\color{darkred}\pmb{\xmark}} & {\color{darkgreen}\pmb{\cmark}} & {\color{darkgreen}\pmb{\cmark}} \\
Groq \cite{groq}                                      & {\color{darkred}\pmb{\xmark}} & {\color{darkred}\pmb{\xmark}} & {\color{darkred}\pmb{\xmark}} & {\color{darkgreen}\pmb{\cmark}} \\
Tesla Dojo \cite{groq}                                & {\color{darkgreen}\pmb{\cmark}} & {\color{darkgreen}\pmb{\cmark}} & {\color{darkred}\pmb{\xmark}} & {\color{mustard}\pmb{\cmark}} \\
Esperanto \cite{esperanto}                            & {\color{darkred}\pmb{\xmark}} & {\color{darkred}\pmb{\xmark}} & {\color{darkred}\pmb{\xmark}} & {\color{darkgreen}\pmb{\cmark}} \\ 
Google TPU~\cite{tpu_google}                          & {\color{mustard}\pmb{\cmark}} & {\color{darkred}\pmb{\xmark}} & {\color{darkgreen}\pmb{\cmark}} & {\color{darkred}\pmb{\xmark}} \\ 
Nvidia A100~\cite{nvidia_a100}                        & {\color{darkgreen}\pmb{\cmark}} & {\color{darkred}\pmb{\xmark}} & {\color{darkred}\pmb{\xmark}} & {\color{darkgreen}\pmb{\cmark}} \\ 
\midrule
\proj                                             & {\color{darkgreen}\pmb{\cmark}} & {\color{darkgreen}\pmb{\cmark}} & {\color{darkgreen}\pmb{\cmark}} & {\color{darkgreen}\pmb{\cmark}} \\
\bottomrule
\end{tabularx}
\vspace{-2mm}
\label{table:related}
\end{table}

\subsection{Chiplets and Fabrication Cost}\label{sec:background_nre}

The semiconductor industry is increasingly utilizing multiple dies in a package~\cite{ibm_telum,shapphire_rapids,ponte_vecchio,amd_rome,amd_vcache} not only to increase silicon lithography yield but also to enable reusing components across different chip products.
From the silicon manufacturing perspective, the Non-Recurring-Engineering (NRE) cost of a wafer mask is so high with respect to the silicon wafer itself that the cost per wafer is 18$\times$ larger when manufacturing 100 wafers, instead of 100,000~\cite{lithovision}.
Since the volume of silicon production of each chip architecture generation is not as high in the HPC market as in the consumer electronic market, we argue that the mass fabrication of a chiplet that serves as a building block for scalable architectures is a valuable proposition.
Moreover, from the supply-chain~\cite{supply_chain} and the embodied-carbon perspective~\cite{chasing_carbon,carbon_explorer,eco_chip}, it is more agile and sustainable to integrate existing chiplets differently to meet the demands of various products than fabricating a new chip for each one.
Thus, we set out to design a chiplet that can be fabricated at high volume to amortize NRE costs, and then integrated into a Multi-Chip Module (MCM) package at the scale and configuration that best meets the performance, power, and cost requirements of the target product, within the irregular application domain.
Beyond being able to create chips for different markets (e.g., single-die for automobile chips or full nodes for HPC clusters), graph-search-oriented supercomputers could integrate larger or smaller chip packages per node, depending on the target performance and cost.

\vspace{-1mm}
\subsection{Design Space of Manycore Systems}\label{sec:background_manycore}

In recent years, the widespread demand for large deep-learning models has accelerated the development of manycore and dataflow systems that can massively parallelize compute-intensive workloads.
Although the demand for graph and sparse linear algebra workloads, as well as target data sizes, are growing, we have not seen systems exhibiting as high scalability as dense systems.
This is because the irregular data-access patterns and the low arithmetic intensity (operations/byte) challenge the network and the memory systems optimized for dense workloads.

Table~\ref{table:related} shows a selection of the recent manycore systems starting with those focusing on sparse workloads and continuing with those focusing on AI.
While some of these manycores have good attributes for sparse data processing (large SRAM capacity~\cite{dalorex,cerebras_fft,groq,graphcore,tesla_dojo} or on-chip DRAM~\cite{nvidia_a100,sambanova,tpu_google,manticore,tesseract}), the ratio between on-chip memory and PUs is optimized for dense computation, and the interconnect is designed for dataflow communication.
And precisely these resources---the network-on-chip (NoC) and the on-chip memory capacity and bandwidth---are the most important ones for irregular applications.
Note that we use chip and package interchangeably, and chiplet to refer to a die that can be integrated into a package with others.

\textbf{\textit{Memory Configurability:}}
Based on our experiments, there is no universally optimal memory-hierarchy configuration for irregular applications that optimizes for all important target metrics (e.g., time-to-solution, and performance per cost or power unit). 
This is exacerbated by the fact that graphs and sparse datasets for different application domains can have very different degree distributions, local structure and size and therefore may present different optimization landscapes.
Therefore, we argue that---given the NRE costs---it is more cost-efficient to design a chiplet architecture that allows for optimizations during packaging than to fabricate a new chip for each configuration.
\cref{fig:cake} illustrates that \proj allows opting whether to integrate DRAM on the package, and it enables interleaving DRAM devices between columns of \device{} dies---not just on the edges of the package as in existing chiplet-based designs.
\cref{sec:approach} describes how \proj allows for this flexibility, which can enable different cost-optimal design points given the variance in HBM prices with availability~\cite{hbm_price}.

\textbf{\textit{Network Configurability:}}
One of the key attributes that improved the performance of irregular applications in Dalorex was using a torus NoC, which helped to reduce network congestion caused by irregular communication patterns.
A torus NoC can be implemented without a long wrap-around wire by \textit{folding} it into the 2D space, i.e., having even-indexed tiles connected, representing one bisection of the network and odd-indexed tiles representing the other bisection~\cite{topology_study}.
However, this sets the size of the torus NoC at silicon time.
Since we want to support integrating a scaling number of chiplets and keep having them connected in a torus, we propose a \textit{reconfigurable} torus NoC where the topology is defined prior to execution (see 
\cref{sec:noc_config}).

\textbf{\textit{Compute Configurability:}}
The NoC and memory configurations are geared towards being able to feed the PEs with data, which is much more challenging for irregular applications than for dense ones.
The number of PEs per chip and per node will ultimately shape the topology of the cluster-level network, the size of the dataset that can be processed, and at what speed.
Thus, being able to defer this decision to post-silicon, is another motivation for having this level of configurability.
We also advocate for the flexibility of ISA-programmable PEs over the efficiency gains of ASICs or CGRAs because the compute area and power needed for irregular applications is low in comparison with the memory and network resources---as we show in \cref{sec:results}.

\section{The \proj Architecture}\label{sec:approach}

Having motivated the need for a reconfigurable architecture for irregular applications to meet different PPA and cost targets, this section describes how \proj architecture achieves that.
First, \cref{sec:noc_config} describes how our Torus NoC design can be configured to span multiple dies and packages.
Then, \cref{sec:sram_config} describes how we manage the tile's SRAM as a cache and/or as a scratchpad, and provide prefetching support by leveraging the task-based execution model.
\cref{sec:reconfig_dram} describes how the DRAM dies can be interleaved in between the \device{} dies.
Finally, Table~\ref{table:configuration_knobs} summarizes the configurable knobs and serves as an outline for our evaluations in \cref{sec:results}.

\textbf{\textit{Dalorex's task-based execution model:}}
We use the latest work on sparse architectures, Dalorex~\cite{dalorex}, as a starting point to build our base design.
We take their task-based execution model, enabled by the Task-scheduling Unit (TSU), which is the nexus between the NoC router and the Processing Unit (PU) executing the tasks.
When the PU executes a task it can potentially spawn new tasks by writing task-invocation parameters into the output queues (OQs)---one per task type.
The router eventually takes them from the OQs and places them into the NoC.
Task invocations are routed to the tile that owns the data they operate on because the dataset layout into the partitioned global address space (PGAS) is statically known.
At the destination tile, the router places a task-invocation message into the tile's input queue (IQ), so that the TSU can schedule the task to the PU.
Since there is one IQ per type of task, the TSU also prioritizes the tasks based on their type, with a heuristic informed by the occupancy of the task queues.

This execution model---where every tile only operates on local data---shifts the problem of irregular memory-access patterns to irregular inter-tile communication.
Dalorex observed that when dealing with irregular traffic a 2D torus provides a more uniform utilization than a 2D mesh~\cite{dalorex}.
To make implementation practical on 2D silicon, the 2D torus is folded (as described in \cref{sec:background_manycore}).
However, Dalorex assumes a monolithic wafer-scale design and does not solve the problem of setting up a 2D torus spanning different grid sizes.

\subsection{Designing a reconfigurable 2D Torus}\label{sec:noc_config}

\begin{figure*}[t]
\centering  
\includegraphics[width=\textwidth]{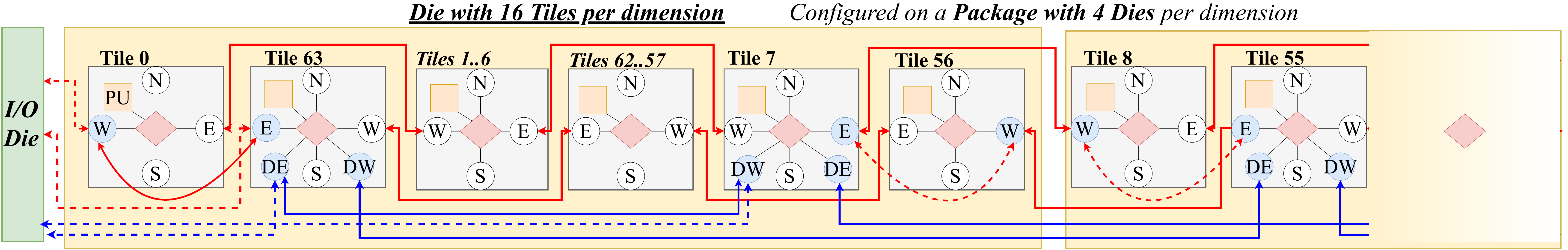}
\vspace{-6mm}
\caption{
Horizontal links within a \device{} die and across dies.
The \textcolor{red}{red links} show the NoC that connects every tile (\textcolor{red}{\textit{tile-NoC}}), while the \textcolor{blue}{blue links} show the NoC that connects to one tile per die (\textcolor{blue}{\textit{die-NoC}}).
Because of the die-NoC, the routers at the die edges are radix-9, while the rest are radix-5.
The ports shadowed in blue are runtime reconfigurable; any tile subgrid within a node board may become torus (including across packages).
The dies on the edges of a package will interface with the \textcolor{darkgreen}{I/O die}.
All I/O links are configured when loading the dataset to maximize I/O bandwidth.
During program execution, both NoCs may become torus, or tile-NoC torus and die-NoC mesh, to keep streaming from I/O.
Note that since the torus is folded, all the links within each NoC are nearly the same length.
Even the longest wires coming from die-NoC are shorter than the 25mm die-to-die limit~\cite{bow} for the integrations shown in \cref{fig:cake}.
}
\vspace{-3mm}
\label{fig:torus}
\end{figure*}

Motivated by our need to scale irregular communication while providing a chiplet-based integration, we design a Torus NoC whose topology can be configured in software.
For example, we can confine the Torus within a die, or have it span multiple dies and packages within the board of a compute node.
(Table~\ref{table:wire_param} shows the interconnect energy and latency assumed for our evaluation for hops at each level.)
\cref{fig:torus} shows that the routers at the edges of each die can be configured to connect to a router on the next die or wrap around by connecting to the adjacent tile (Tiles 0 and 63).
Moreover, we can reconfigure a 2D-torus NoC into two 2D-mesh NoCs by not connecting the wrap-around links on the routers at the edges.
This allows for a 2D-mesh NoC to be used for streaming data from I/O to the tiles, and then reconfigure it to a 2D-Torus NoC for the rest of the execution.

\textbf{\textit{Reducing long-distance communication:}}
To reduce the number of hops across chiplets, \proj has two hierarchical NoCs, one that connects every tile and one that hops once per die, as shown in \cref{fig:torus}.
Each NoC topology is individually configured.
While bringing the data inside the package from the disk, they both would be configured as a mesh to increase I/O.
During the execution, both may become torus, or the die-NoC may also remain open to stream I/O data.

After doing a literature review, we found that---to the best of our knowledge---our proposal is the first 2D-Torus NoC topology for which the size of the network can be configured at runtime, being able to span one torus across multiple dies, or have multiple torus networks.

\textbf{\textit{Off-Chip Bandwidth:}}
\proj employs I/O chiplets to connect packages.
This allows \device{} dies to remain agnostic to a specific off-chip protocol (e.g., PCIe 6.0~\cite{pcie6}).
It also defers choices like off-chip bandwidth to packaging time, to decide based on the requirements of the final product.
The off-chip bandwidth could be as high as the I/O-\device{} bandwidth.

\subsection{Reconfigurable SRAM Management}\label{sec:sram_config}

In Dalorex, all the dataset arrays had to fit within the aggregated SRAM memory of the grid of tiles.
While this is suited for large levels of parallelization that aim to achieve the fastest time-to-solution, being forced to scale out with the dataset size can become very costly (shown in \cref{sec:res_packaging_time}).
Thus, we need to support data caches that are backed up by DRAM.

Keeping the same tile design as in Dalorex, we manage the SRAM scratchpad vastly differently.
Inspired by prior work on reconfigurable caches~\cite{amoeba,cameo,esp_cache}, \proj offers to the software the appearance of a tile's local address space, which is mapped to the SRAM as a cache and/or as a scratchpad.

\textit{\textbf{Cache mode:}}
The cache mode allocates a portion of the SRAM as a direct-mapped cache, for a given address range of the tile's local address space (a segment).
It stores cacheline tags and the valid/dirty bits in SRAM too, so the area overhead of the cache is only the logic for tag comparison.
To minimize the hardware and energy overhead of using the SRAM as a cache, we make the caches directly mapped.
The cacheline width equals the bitline width of the DRAM memory controller (512 bits in our experiments).

\textit{\textbf{Scratchpad mode:}}
When scaling out the parallelization of a dataset, if the memory footprint per tile fits in the local SRAM, the data cache would not be configured.
Instead, the data segment would map directly into the SRAM scratchpad.
A data segment can also be mapped directly to the SRAM scratchpad, even when the data cache is enabled.

\textbf{\textit{Data cache misses and evictions:}}
Upon a miss, the data cache fetches the full cacheline from DRAM without checking for coherence since the data arrays in the data segment are not shared.
This fetch uses a physical address since each DRAM module is private to each \device{} die with 1-1 mapping between tiles and DRAM address space.
Each tile's local SRAM is backed up by a DRAM storage of size $DRAM\_capacity / tiles\_per\_die$.
(The details of the DRAM devices assumed for the evaluation are described in Table~\ref{table:wire_param}.)
The data cache has a dirty bit per line to write back to DRAM upon eviction.
Since the data cache of each tile only contains the part of the dataset that the tile is responsible for, there are no coherence issues for modified data.

\textbf{\textit{Prefetching:}}\label{sec:prefetching}
The PU has a very simple in-order pipeline, which stalls waiting for data on a D\$ miss.
Since the first parameter of every task message contains an array index, and the TSU knows to which array it corresponds (used for NoC routing), we use this information to prefetch the data.
A task may access more than one array using the same index, so we add another pointer to the TSU's per-task table so that it prefetches both when needed.
Those tables also have an extra bit saying whether the PU keeps prefetching data during the task execution.
Since pointer indirections are split into tasks, when tasks access multiple array elements, they often do it with a streaming pattern, 
To prefetch these, we enable PU's next-line prefetcher for tasks that access multiple elements.

\subsection{Reconfigurable On-chip Memory Capacity}\label{sec:reconfig_dram}

\begin{figure}[t]
\centering
\vspace{-1mm}
\includegraphics[width=\columnwidth]{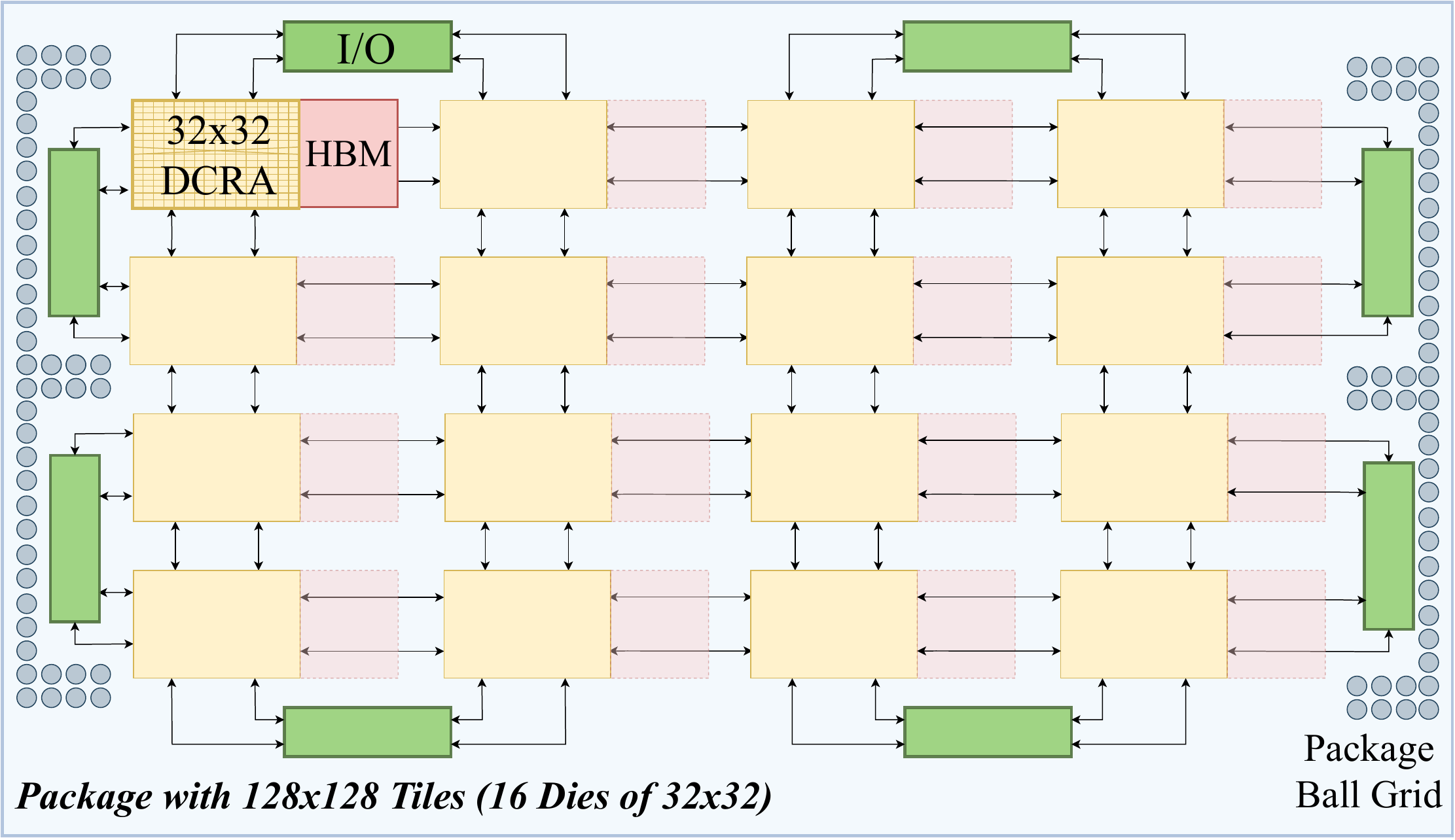}
\vspace{-3mm}
\caption{
Top view of a package with 128x128 tiles.
The number of \device{} dies would determine the compute capacity, while including HBM dies would determine the compute-to-memory ratio of the package.
The number of I/O dies (and their bandwidth) determines the off-chip bandwidth.
}
\vspace{-5mm}
\label{fig:package}
\end{figure}

The memory hierarchy determines the bandwidth available to computing units and, thus, at which level of arithmetic intensity the chip becomes memory-bound.
To increase memory bandwidth, prior work has proposed having DRAM on the chip.
Prior work proposed 3D integrations where processing cores are on the base plane of the 3D-stacked memory, accessing the TSVs directly~\cite{tesseract,2018graphp,2019graphq,pim_hbm}.
However, this may render impractical.
There are very few foundries that can fabricate HBM technology, and the design seems to only change with every generation of JEDEC specification~\cite{jedec_2015} (e.g., HBM2, HBM2e, HBM3) which vendors conform to.
So it would be expensive---if possible---to fabricate an HBM die with a custom base layer.
Moreover, the base layer of the HBM stacks is already filled with MBIST logic and PHY~\cite{hbm2_sk_hynix,hbm3_sk_hynix,hbm,hbm_samsung}.
Alternatively, other works have proposed integrations of off-the-shelf HBM dies with compute dies by having the compute dies surrounded by HBM devices on the edges of the chip~\cite{amd_epyc_isca,amd_rome,emib,kaby_lake,shapphire_rapids,manticore,nvidia_chiplets,nvidia_a100}.
\proj is the first proposal for having horizontally integrated HBM dies (2.5D) interleaved across columns of compute dies (as shown in \cref{fig:package}).
This allows the design of \device{} dies to be agnostic to the number of \device{} and DRAM dies that are eventually integrated on-chip.

\textbf{\textit{Interleaving DRAM \& \device{} dies:}}
Figs.\ref{fig:cake} and \ref{fig:package} depict the integration of \device{}, and HBM dies via a passive silicon interposer.
This provides higher \device{}-HBM bandwidth than a silicon bridge~\cite{emib} or substrate integration~\cite{flipchip_options,amd_rome}.
The interposer only contains the wires between a \device{} die, and its HBM die since HBM access is exclusive.
The links connecting \device{} dies are routed through the organic substrate, which also contains the power delivery and redistribution layer.

\textbf{\textit{Memory controller:}}
As depicted in \cref{fig:cake}, the memory controller lives on the \device{} die, and thus, in case of not integrating DRAM, it becomes \textit{dark silicon}. Although in our evaluation, we consider it as such, future work could investigate using an embedded FPGA logic that allows hardening a memory controller of choice (for HBM or DDR, opting for high bandwidth or high capacity) or other computing logic (in the case DRAM is not integrated).

\subsection{Choosing the Right Configuration}\label{sec:config_table}
Now that we have established the design of the \proj architecture, we are ready to discuss the design space exploration that we performed in our evaluation to find the adequate pre-silicon options for the sparse applications that we target.
Table~\ref{table:configuration_knobs} shows the pre-silicon options, and post-silicon and compile-time configurations that can be selected for the \proj architecture.
In \cref{sec:results} we present our evaluation results for most of these features, in the order that they are presented in \cref{table:configuration_knobs}.
Moreover, in \cref{fig:flowchart} we provide guidance on how to choose the pre-silicon and packaging-time configuration of \proj based on the target deployment of the final product.

\begin{scriptsize}
\setlength{\tabcolsep}{1pt}
\begin{table}[t]
\caption{Reconfigurable Parameters of \proj}
\centering
\small
\begin{tabularx}{\columnwidth}{l}
\toprule
\textbf{Tapeout-time Design Decisions} \\
\midrule
1. \# of Tiles per die \\
2. \# of PUs per tile (and their operating and max. frequency) \\
3. SRAM capacity per tile (KB) \\
4. NoC Width (and the operating and max. frequency)\\
\midrule
\textbf{Packaging-time Design Decisions} \\
\midrule
5. \# of \device{} dies per package\\
6. \# of DRAM dies per package and capacity of each (GB)\\
7. \# of I/O dies per package (Off-Package BW) \\
\midrule
\textbf{Compile Time Configurations} \\
\midrule
8. Size of the input and output queues (for every task type) \\
9. Size and Place of the grid that the workload uses (grid of dies)\\
10. The address space for which the data is cached \\
11. Size of the D\$ (in data elements) \\
\bottomrule
\end{tabularx}
\label{table:configuration_knobs}
\vspace{-3mm}
\end{table}
\end{scriptsize}

\section{Evaluation Framework}\label{sec:methodology}

This section describes the methodology we use to evaluate \proj, starting with the applications and datasets we use, followed by the simulation infrastructure we built, including the cost model we use to evaluate the cost-effectiveness of different configurations.

\subsection{Applications and Datasets}\label{sec:eval_applications}

We evaluate the performance of \proj on four graph workloads, SPMV, and histogram~\cite{parboil}.
\textit{Breadth-First Search (BFS)} determines the number of hops from a root vertex to all vertices reachable from it;
\textit{Single-Source Shortest Path (SSSP)} finds the shortest path from the root to each reachable vertex;
\textit{PageRank} ranks websites based on the potential flow of users to each page~\cite{pagerank};
\textit{Weakly Connected Components (WCC)} finds and labels each set of vertices reachable from one to all others in at least one direction (using graph coloring~\cite{connected_components});
\textit{Sparse Matrix-Vector Multiplication (SPMV)} multiplies a sparse matrix with a dense vector.
\textit{Histogram} counts the values that fall within a series of intervals.
We report traversed edges per second as $TEPS=m/time$ where $m$ is the number of edges connected to the vertices in the graph starting from the search key.
When we report TEPS for SPMV and Histogram, we consider the non-zero elements to multiply, and elements to process, respectively.

We use three sizes of the RMAT~\cite{kron} graphs---standard on the Graph500 list~\cite{graph500}---RMAT-22, RMAT-25 and RMAT-26, which are named after their number of vertices.
For example, RMAT-26 (abbreviated as R26 in \cref{sec:results}) contains $2^{26}$, i.e., 67M vertices (V) and 1.3B edges (E), and has a memory footprint of 12GB.
We also use the Wikipedia (WK) graph (V=4.2M, E=101M) in our evaluation to exercise different graph topologies.
For SPMV we use the same datasets as a graph can be seen as a square sparse matrix with V rows and columns and E non-zero elements.
The graphs (as sparse matrices) are stored in the Compressed Sparse Row (CSR) format without any partitioning, i.e, the dataset contains three input arrays, one for the values of the non-zeros, one for the column indices of those non-zeros, and one for the pointers to the beginning of each row in the previous two arrays.
The output array has size V, and its meaning depends on the application, e.g., for Histogram, it is the count of the column indices of the non-zeros.

\subsection{Measuring Performance and Energy}\label{sec:perf_simulation}
We built our evaluation framework on top of the functional simulator from Dalorex~\cite{dalorex}---a cycle-accurate simulator for the NoC and based on performance models for the PUs---which we vastly extended to support chiplets, caches and DRAM.
We chose this approach because faithfully modeling the NoC and memory hierarchy is the most critical part for these large parallelizations of data-dependant applications.

The simulation of the PUs is done by instrumenting the application code with the delay of each instruction.
In this paper, we make the same assumption as in Dalorex, of one instruction per cycle.
However, while Dalorex assumed also one cycle for memory operations, we introduced a detailed model for the memory hierarchy.

Table~\ref{table:wire_param} summarizes some of our additions to the energy, latency, and area model for communication links and memory technology.
The full set of energy, area and performance parameters we used for our evaluation can be found in our \repository, which also includes a Readme file will all the scripts used to generate the artifacts and plot the evaluation figures in this paper.

Since we test different frequencies for the NoC and the PUs on Figs.~\ref{fig:NoC_characterization} and \ref{fig:PU_freq}, our area and energy numbers scale with frequency and voltage.
For the rest of the experiments, our operating frequency is 1Ghz.

\subsection{Cost Model}\label{sec:perf_simulation}
We also added a cost model to the simulator to study the cost-effectiveness of different \proj configurations.
The cost model---similarly to the energy model---is decoupled from the runtime simulation process, i.e., cost and energy can be re-calculated post-simulation for different parameters. This is useful to study how price variations can change the cost-effectiveness of different configurations.

\textbf{\textit{Silicon cost:}}
All of our evaluations consider 7nm technology for the \device{} chiplets; we assume that a 300mm wafer with this transistor process costs \$6,047~\cite{lithovision}.
We obtain the cost per die by dividing the wafer cost by the number of good dies, which we calculate using 0.2mm scribes, 4mm edge loss, and 0.07 defects per $mm^2$.
(We integrate and validate~\cite{yield_calc} die yield calculations in our cost model using Murphy's model.)
When comparing the cost-effectiveness of our results, we do not include the Non-Recurring Engineering (NRE) cost of the \device{} dies since all the options use the same technology.

\textbf{\textit{Packaging cost:}}
In terms of packaging, all our results featuring grid sizes over 128x128 use multiple packages (of 64x64 tiles each, based on our rationale from \cref{sec:config_table}).
We assume the cost of the 65nm silicon interposer connecting a \device{} die with HBM (including bonding) to be 20\% of the price of a \device{} die~\cite{cost_model_package}; the cost of an organic substrate to be 10\% of the price of an equal-sized \device{} die, and the bonding to add an additional 5\% overhead~\cite{cost_model_interposer,bonding_yield}.

\textbf{\textit{DRAM cost:}}
Regarding \textit{DRAM}, we assume an 8GB HBM2E device with eight 64GB/s memory channels.
While this cost is not disclosed, we made an educated guess using public sources~\cite{lithovision,cost_hbm}.
We assume 7.5\$/GB, which is more affordable than when HBM was released in 2017.
One could expect this price to decrease over time as more vendors fabricate HBM or the process matures~\cite{micron_hbm,hbm2_sk_hynix,hbm_samsung,hbm3_sk_hynix}.

\begin{table}[t]
\centering
\small
\caption{Energy (E), bandwidth, latency, and area of links and memory devices assumed for the evaluation.}
\begin{tabularx}{\columnwidth}{@{\hspace{0pt}}l@{\hspace{-25pt}}r@{\hspace{0pt}}}
\toprule
\textbf{Memory Model Parameters} & \textbf{Values} \\
\midrule
SRAM Density\ & 3.5 MB/mm$^2$~\cite{renesas_ff7nm} \\
SRAM R/W Latency \& E. & 0.82ns \& 0.18 / 0.28 pJ/bit~\cite{renesas_ff7nm} \\
Cache Tag Read \& cmp. E. & 6.3 pJ~\cite{renesas_ff7nm,ariane_cost}\\
HBM2E 4-high Density\ & 8GB/110mm$^2$ (75 MB/mm$^2$)~\cite{hbm2_sk_hynix} \\
Mem.Channels \& Bandwidth & 8 x 64 GB/s~\cite{hbm2_sk_hynix} \\
Mem.Ctrl-to-HBM RW Latency \& E. & 50ns \& 3.7pJ/bit~\cite{fine_grain_dram,hbm_samsung_power} \\
Bitline Refresh Period \& E. & 32ms \& 0.22pJ/bit~\cite{refresh_time_hbm,dram_activation_energy} \\
\midrule
\textbf{Wire \& Link Model Parameters}  & \textbf{Values} \\
\midrule
MCM PHY Areal Density                   & 690 Gbits/mm$^2$~\cite{die2die_comp} \\
MCM PHY Beachfront Density              & 880 Gbits/mm~\cite{die2die_comp} \\
Si. Interposer PHY Areal Density        & 1070 Gbits/mm$^2$~\cite{die2die_comp} \\
Si. Interposer PHY Beachfront Density   & 1780 Gbits/mm~\cite{die2die_comp} \\
Die-to-Die Link Latency \& E.          & 4ns \& 0.55pJ/bit ($<$25mm)~\cite{bow} \\
NoC Wire Latency \& E.              & 50 ps/mm \& 0.15pJ/bit/mm~\cite{pim_hbm} \\
NoC Router Latency \& E.            & 500ps \& 0.1pJ/bit \\
I/O Die RX-TX Latency                   & 20ns~\cite{pcie6} \\
Off-Package Link E.                 & 1.17pJ/bit (upto 80mm)~\cite{nvidia_chiplets}\\
\bottomrule
\end{tabularx}
\vspace{-2mm}
\label{table:wire_param}
\end{table}

\vspace{-1mm}
\section{Results}\label{sec:results}

Unlike dense-data workloads, data traversal algorithms have highly varying demands that depend on data size and structure (e.g., skewness), making the peak compute performance of current systems severely under-utilized.
\proj is an architecture for efficiently processing data traversal workloads with several pre-silicon options as well as packaging- and compile-time configurations that make this process cost-efficient.

We start in \cref{sec:res_default} by demonstrating the benefits of the default options of \proj such as the utilization of torus topology within and across chiplets.
\cref{sec:res_target_app} then analyze pre-silicon options, such as SRAM size per tile and PU frequency, where we can make tradeoffs depending on the target set of application domains.
Next, \cref{sec:res_skewness} analyzes pre-silicon options where decisions may be made depending on the skewness of dataset sparsity, such as PUs per tile, tiles per chiplet, and NoC frequency.
In \cref{sec:res_packaging_time} we evaluate packaging-time configurations such as the target-metric-dependent choice of including HBM or not on the package. 
In \cref{sec:million} we evaluate compile-time configurations,
\ifarxiv
like the sizes of the tiles' input and output queues, and how the level of parallelization for a given dataset affects throughput per dollar and watt.
\else
like the level of parallelization for a given dataset, to see how it affects throughput per dollar and watt.
\fi

\begin{figure}[htp]
\includegraphics[width=\columnwidth]{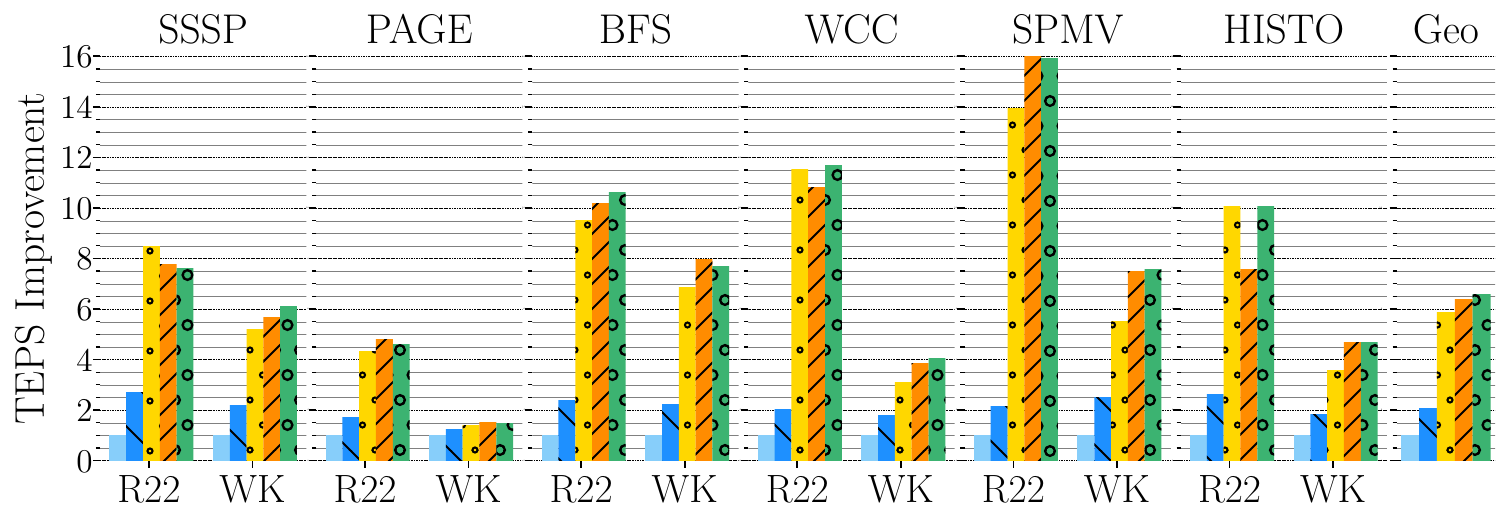}
\includegraphics[width=\columnwidth]{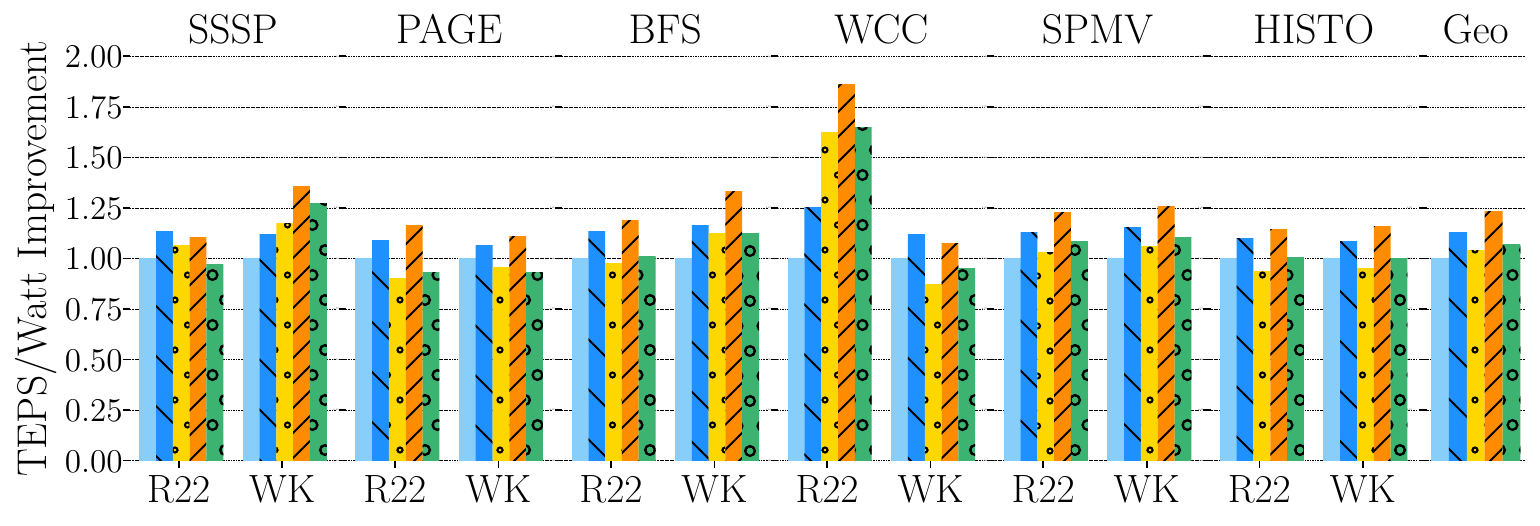}
\includegraphics[width=\columnwidth]{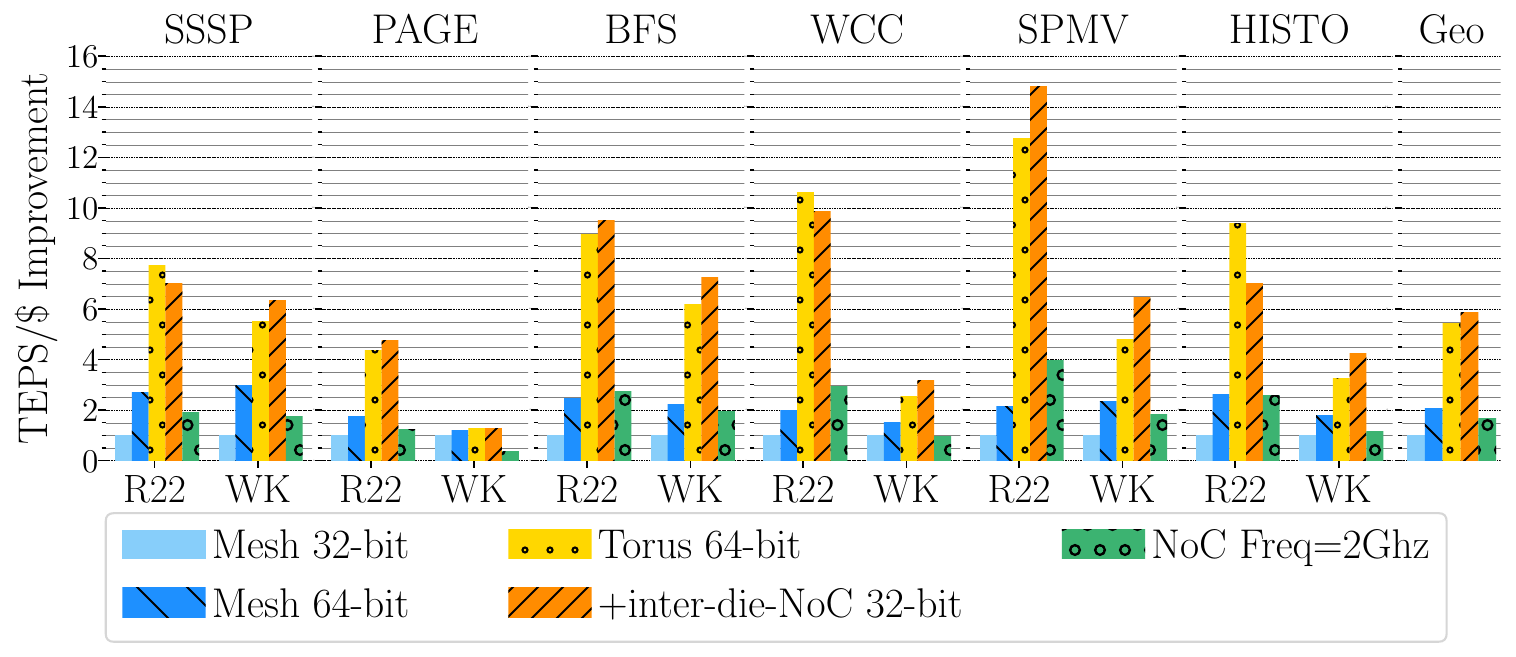}
\vspace{-5mm}
\caption{Performance, energy efficiency, and performance per dollar improvements of different network choices over a baseline of a 32-bit 2D-Mesh.
All configurations use 64x64 tiles (across 16 chiplets), with 512KB SRAM/tile.
}
\label{fig:NoC_characterization}
\vspace{-3mm}
\end{figure}

\subsection{Default \proj Options}\label{sec:res_default}

\cref{fig:NoC_characterization} shows the performance improvement of \proj with different NoC configurations with 64x64 tiles (across 16 chiplets) with 512KB SRAM per tile.
Since at that parallelization level ($2^{22}$ vertices across $2^{12}$ tiles) the network is the bottleneck, we observe a large performance impact when increasing bisection bandwidth and reducing the diameter of the network.
When doubling the width of the Mesh NoC, the performance also doubles, nearly equally across the board.
The largest impact comes from using a Torus as a topology, $2.6\times$ on geomean, but up to $8\times$ for SPMV.
This is expected since not only does Torus increase the bisection bandwidth and reduce the diameter of the network, but also improves the uniformity of the network traffic.
This is aligned with what was observed by Dalorex, and it was the reason why we put effort into designing our reconfigurable network topology where the size of the folded torus can span one or multiple chiplets.
However, the Torus is more power-hungry than the Mesh, and even though it increases throughput, the energy efficiency is 92\% of the Mesh.
This is overcome by using our 2-level hierarchical Torus topology where there is a torus that connects adjacent tiles (even across chiplets) and another torus that connects to one tile per chiplet (\textit{inter-node-die}).
This addition improves the energy efficiency by 19\% and performance by 9\% on geomean, over the regular Torus.
In terms of cost efficiency, the \proj hierarchical Torus has $5.1\times$ higher throughput-per-dollar than the 32-bit Mesh on geomean, for this set of applications and datasets.

Finally, \cref{fig:NoC_characterization} also tests the performance of increasing the network frequency from 1GHz to 2GHz (as a double-pumped NoC~\cite{double_pumping,polaris}).
After the Torus addition, at this scaling level (64x64), the NoC is not the bottleneck anymore, and so the performance improvement is only 3\% on geomean.
However, it is interesting to show the energy and cost impact of this double-pumped NoC.
The reason why the cost increases three-fold is that we set in the area model that the NoC area doubles for this, but also the area of the IO dies on the package (need to handle twice the bandwidth). 
However, this type of NoC could render useful to parallelize huge graphs (potentially skewed) across larger systems, where the NoC is critical. 

\textbf{\textit{Default:}}
We choose the hierarchical Torus as the default NoC topology for \proj, as it provides the best performance per dollar and energy efficiency.
While the 2Ghz NoC option would only render cost-efficient when the configuration or application benefits from extra bandwidth (e.g. when having more than one PU per tile~\cref{sec:res_skewness}).

\subsection{Target Application Dependent Pre-Silicon Options}\label{sec:res_target_app}

One of the main considerations for production systems is the range of application types a system will be serving.
It is not uncommon for workloads that process sparse data to be processed alongside dense-data processing workloads or kernels.
However, there are also target systems that mostly focus on sparse data processing~\cite{iarpa_agile}.
While these application domains have different arithmetic intensities and data-communication needs, there are several factors like SRAM size and maximum PU frequency where we can decide one option or another depending on whether the target application domain is sparse-only or also dense.

\begin{figure}[t]
\includegraphics[width=\columnwidth]{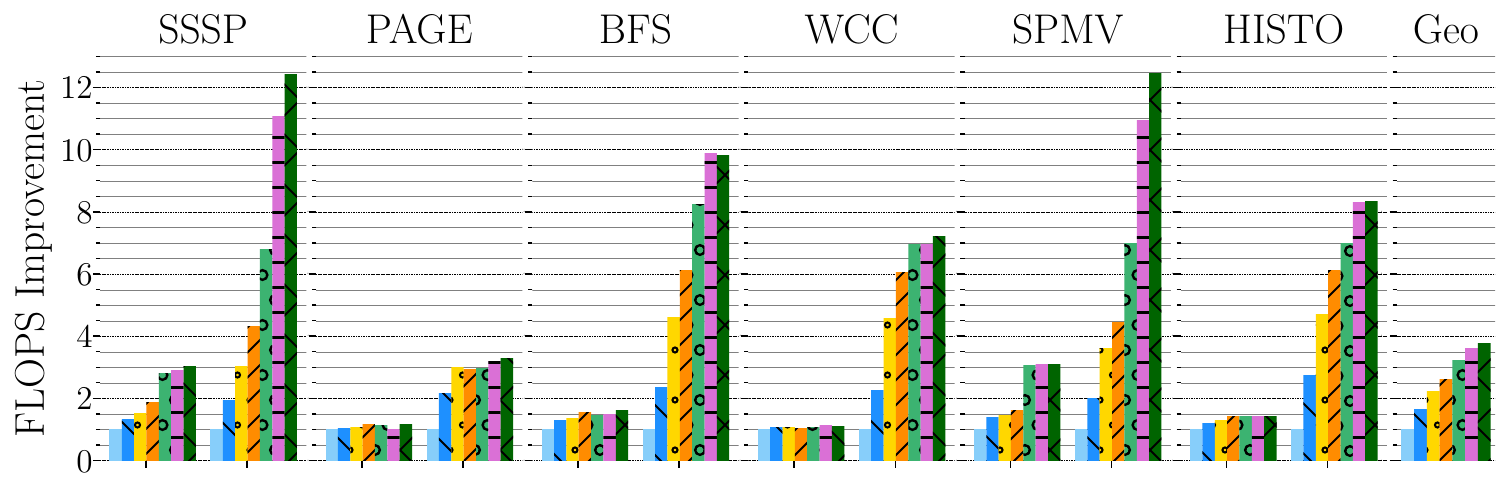}
\includegraphics[width=\columnwidth]{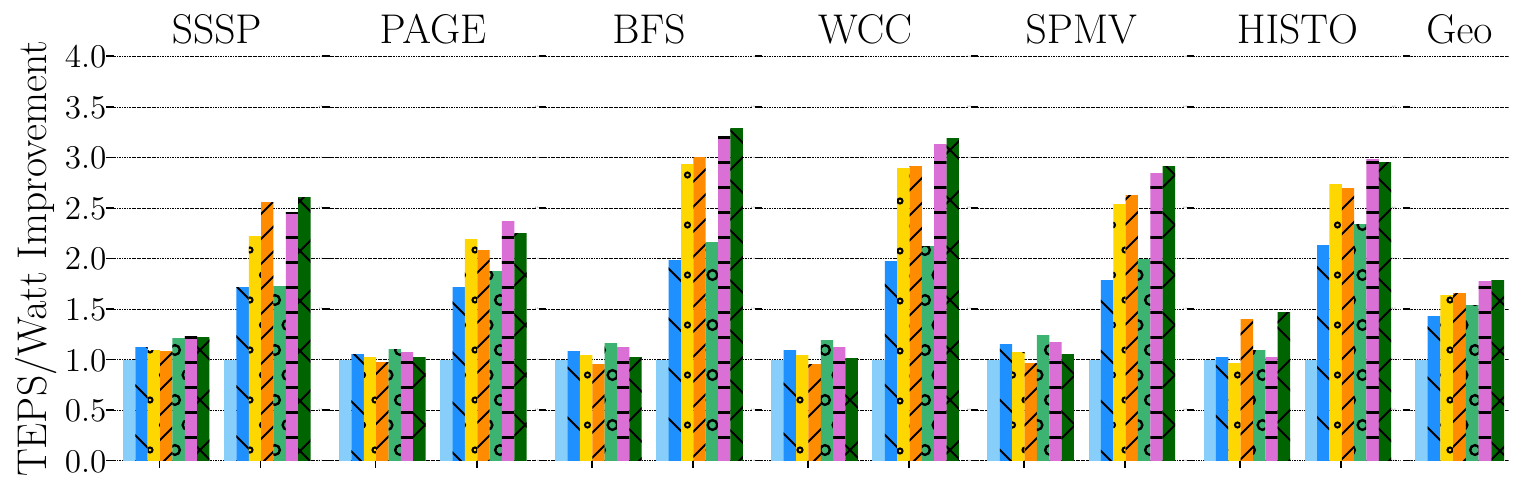}
\includegraphics[width=\columnwidth]{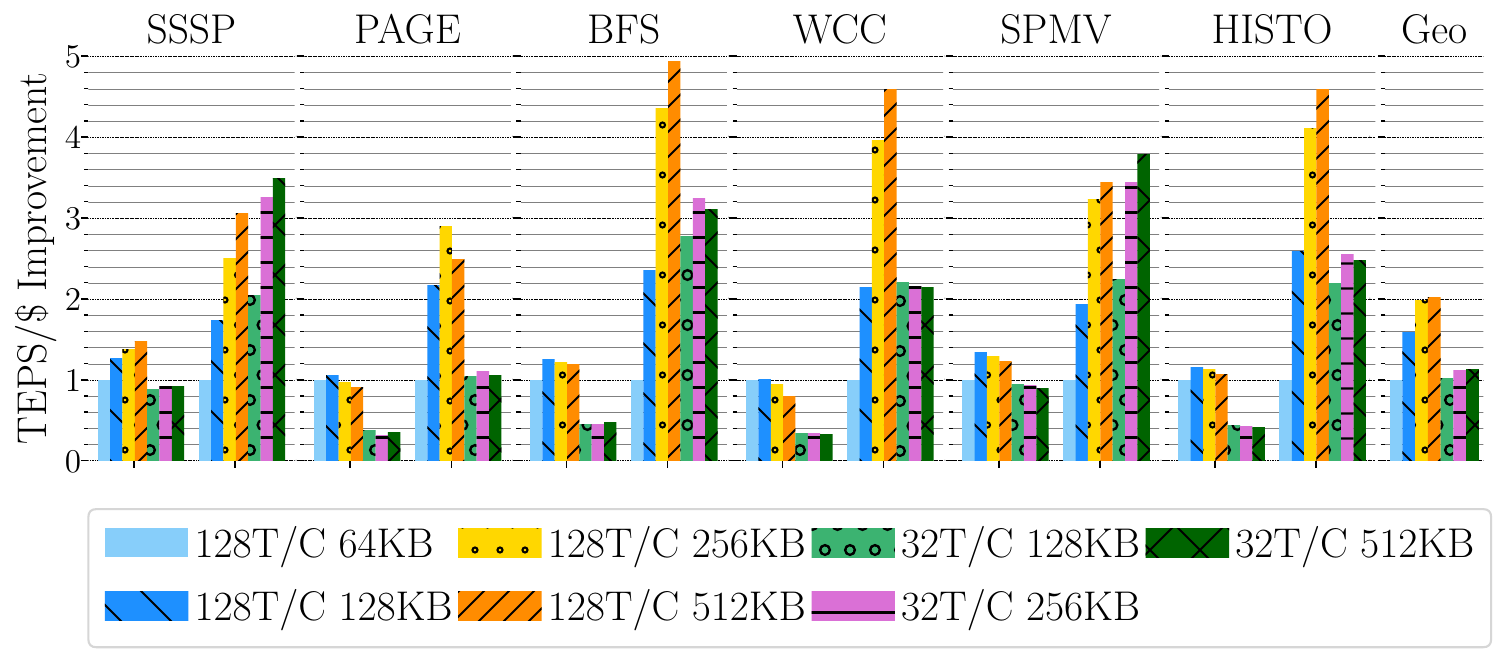}
\vspace{-5mm}
\caption{Performance, energy efficiency, and performance per dollar improvements of different SRAM sizes and number of tiles per HBM channel, over a baseline of 64KB SRAM and 128 tiles per HBM channel (T/C).
A \proj chiplet is always attached to a single 8-channel HBM device, and thus, the number of tiles per chiplet determines the ratio of tiles per HBM channel.
}
\vspace{-5mm}
\label{fig:sram_configurations}
\end{figure}

\cref{fig:sram_configurations} shows the performance, energy efficiency, and performance per dollar improvements of different SRAM sizes and different numbers of tiles per \proj chiplet, but the same number of tiles in total (1024, organized as 32x32).
Since a \proj chiplet is always attached to a single 8-channel HBM device, the number of tiles per chiplet determines the ratio of tiles per HBM channel (T/C).
Since these sparse applications have a low arithmetic intensity, i.e., 1.44, 0.8, 1.8, 0.88, 1.52, 0.8 FLOPs/byte, for SSSP, PageRank, BFS, WCC, SPMV, and Histogram, respectively (nearly identical across these datasets), they require a large amount of memory bandwidth.
As a result, \cref{fig:sram_configurations} shows a strong performance increase with SRAM size, $2.6\times$ on geomean when increasing from 64KB to 512KB, with the same chiplet configuration of 32x32 tiles.
Note that these improvements are on geomean across two datasets sizes.
While the RMAT-22 dataset (shortened as R22) has a data footprint ranging from 512MB for Histogram to 1GB for SPMV (0.5-1MB per tile), R25 has $8\times$ larger footprint.
We evaluated this to see the throughput per dollar difference when the dataset nearly fits on SRAM, and when it does not.
Across applications, SSSP and SPMV, with the largest footprint, have the largest performance improvements.
The case of Pagerank is special, as the barrier-synchronization at each epoch is causing low utilization towards the end of each epoch due to work imbalance (shown in \cref{fig:PU_granularity}).

The hit-rate of the data cache increases from 88\% to 96\% on geomean across datasets (from 81\% to 95\% considering only R25).
Since the effective bandwidth of a tile is: $SRAM\_bandwidth \times hit\_rate + DRAM\_bandwidth \times (1 - hit\_rate)$, when the DRAM-bandwidth per tile is low (as it is shared across 128 tiles), the hit-rate has a large impact on the effective bandwidth.
Changing the number of tiles per chiplet to $16x16$ quadruples the DRAM bandwidth per tile, unleashing further performance improvements, $1.44\times$ on geomean across datasets ($1.7\times$ considering only R25) when keeping the same SRAM size of 512KB.
However, this configuration has nearly half the performance per dollar on geomean, due to the higher cost of having four times more HBM devices.

\textbf{\textit{Default:}}
Having observed these results, we choose the 512KB/tile and 32x32 tiles per chiplet as the default configuration of \proj despite having a third of the performance for SSSP and SPMV for R25, because of three reasons.
(1)~the extra cost of more HBM only renders beneficial when the application is memory-bound and with a large data footprint per tile. However, this footprint decreases as we scale the parallelization, aiming for the fastest time-to-solution. Moreover, applications with higher arithmetic intensity will not benefit as much from this bandwidth, making the per-dollar performance worse.
(2)~having more tiles better amortizes the area of the PHY that connects to other tiles; its size, 255$mm^2$, still achieves a good fabrication yield.
(3)~the SRAM size determines the minimal parallelization required for a given dataset for the chip integrations of \proj that do not include HBM---making 512KB beneficial in that case too.

\textit{\textbf{Targeting sparse-dense:}}
If we were to tape out a different configuration for a deployment that would both use sparse and dense applications, having a 128 or 256KB SRAM would represent a better tradeoff, as 256KB achieves only 18\% worse performance than the 512KB on geomean, and equal energy and cost efficiency.
This 32x32 tile per die and 256KB case is the one depicted in \cref{fig:cake}.

\subsection{Data- or Application-Dependent Pre-Silicon Options}\label{sec:res_skewness}

For systems focusing on sparse data processing only, one important consideration is target-data properties such as skewness of sparsity.
In single-data-owner execution schemes like Dalorex, this can create work-imbalance especially when increasing the parallelization level.
This can be accommodated in tile-based systems as well by having multiple PUs per tile, and having them share an input queue (IQ)~\cite{polygraph}.
That way, the hotspots caused by dataset skewness are softened by a shared work queue across PUs, provided that the memory operations are atomic across the tile's PUs.

\begin{figure}[htp]
\includegraphics[width=\columnwidth]{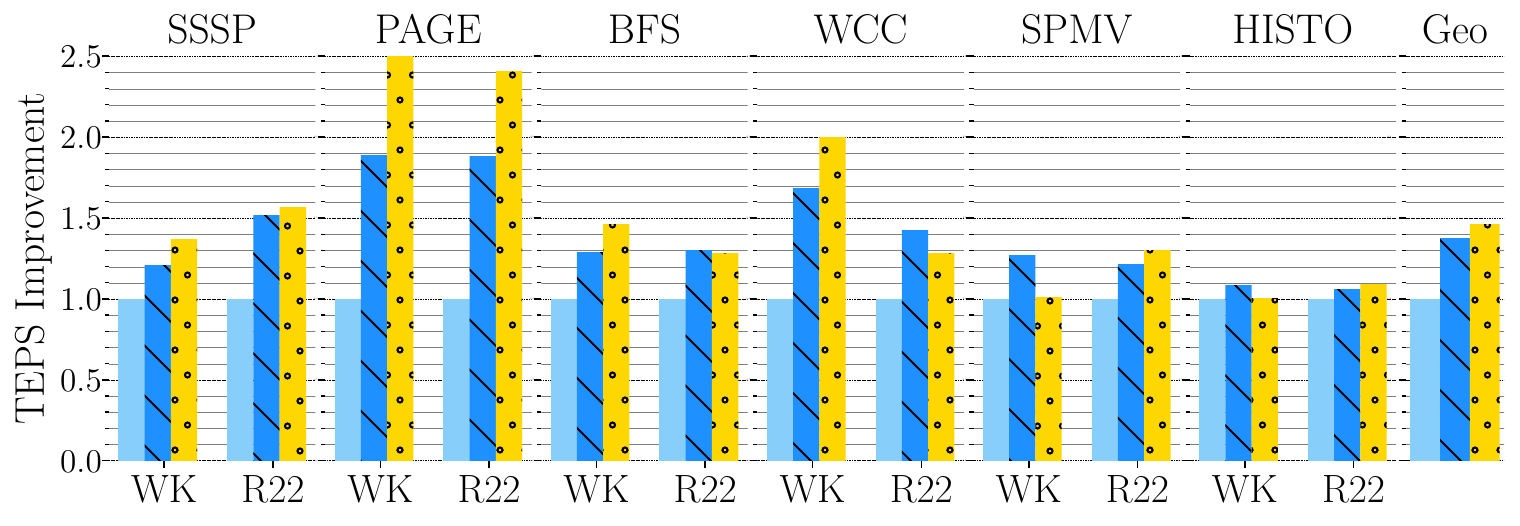}
\includegraphics[width=\columnwidth]{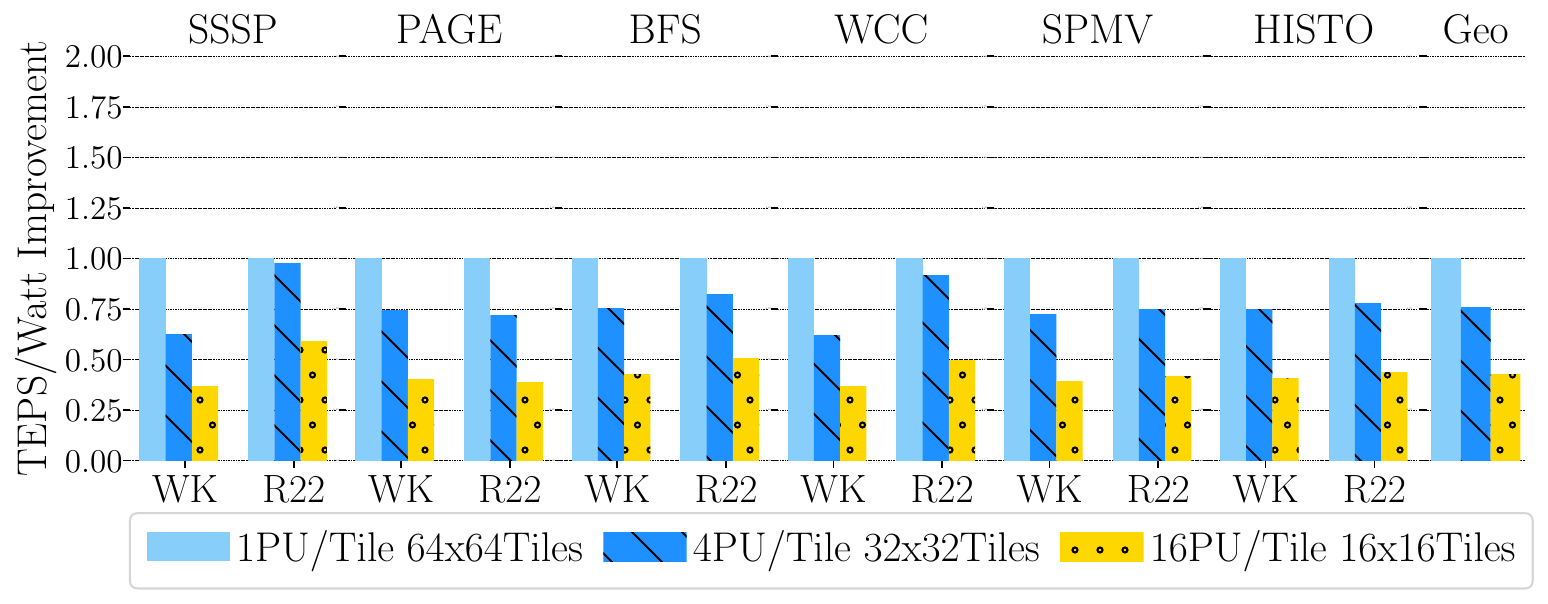}
\vspace{-5mm}
\caption{
Performance and energy efficiency improvements for configurations with 4 and 16 PUs per tile, over the baseline of 1 PU per tile.
All three configurations have 64x64 PUs and the same number of chiplets, but decreasing number of tiles per chiplet, i.e., larger tiles.
To make configurations have the same bisection bandwidth and total SRAM we scale up these resources as the number of tiles is reduced.
}
\vspace{-1mm}
\label{fig:PU_granularity}
\end{figure}

\cref{fig:PU_granularity} evaluates three configurations with similar area and memory and network resources to try to measure only the impact of the tile granularity.
The configuration of 4 PUs per tile has, thus, 4 times fewer tiles but 4 times larger.
These experiments aim to isolate the impact of having more PUs sharing the same IQ, which leads to better load balancing.
(Note that we do not model SRAM bank conflicts on the SRAM but we do consider larger SRAM access latencies for larger SRAM sizes at a rate of one extra nanosecond for each four-fold increase in SRAM size.)  Pagerank benefits the most from work balancing, $2.5\times$ with 16 PUs/tile over the baseline of 1 PU per tile, since this reduces the time wasted at the end of each epoch synchronization.
The other barrier-less graph implementations or single-pass applications (Histogram and SPMV) do not benefit as much.
WCC is the application that experiences the largest differences across datasets, while the other applications have similar performance across datasets.

\textbf{\textit{Default:}}
A single PU per tile is \proj default configuration because despite the benefits of balancing work better across PUs, if we aim to keep the same memory bandwidth per PU, the SRAM size grows too large and so does the distance between the PUs to any given memory bank, increasing both latency and power.
This loss in energy efficiency outperforms the benefits for all barrier-less applications and only renders it interesting for Pagerank.
In general, 4 PUs per tile results in a better balance of performance improvements and energy efficiency than the 16 PUs per tile configuration.

\textit{\textbf{Targeting sparse-dense or synchronization-intensive applications:}}
We could expect 4 PUs per tile to offer a better balance when targeting both sparse and dense applications and having opted for a smaller SRAM size per PU (e.g., 128 or 256KB).
Alternatively, if we expect to focus on synchronization-intensive applications or running skewed datasets, 4 PUs per tile could be a better option.

Another option to adapt the chip to the application domain is supporting different PU frequencies.
Dynamic-voltage frequency scaling (DVFS)~\cite{dvfs} is a common technique to reduce power when the chip is too hot or when we want to save energy.
We argue that it can be also used to adapt for different application domains, but it requires the PU to be designed and synthesized for a target maximum frequency. Thus, this maximum frequency is a pre-silicon option.

\begin{figure}[htp]
\includegraphics[width=\columnwidth]{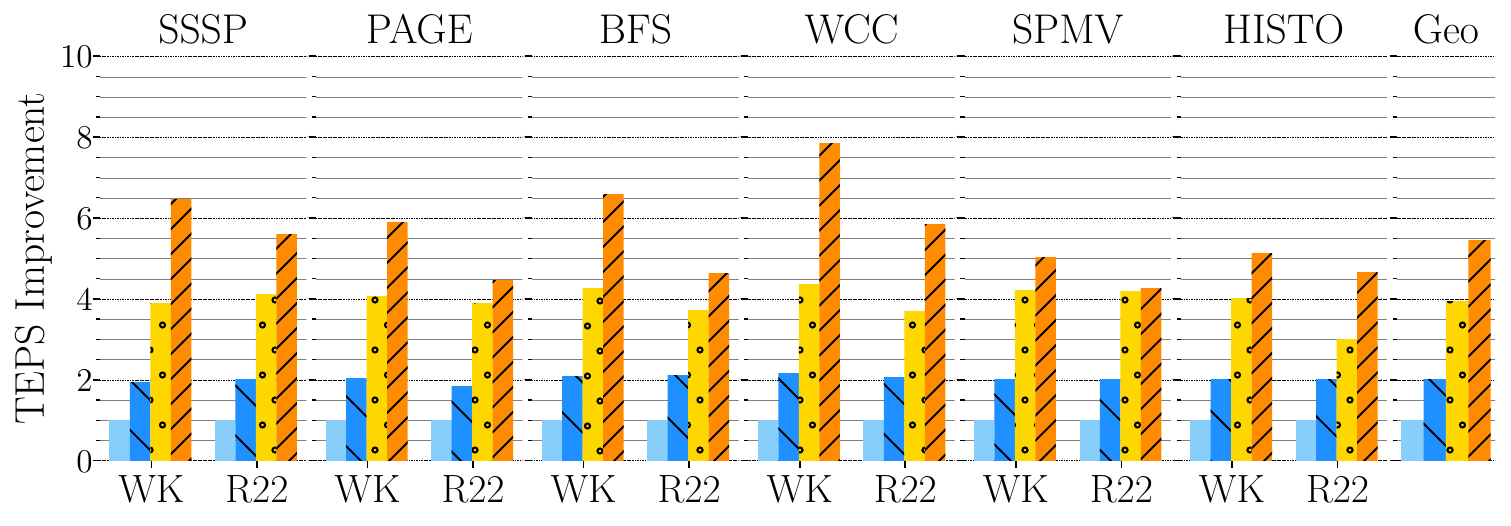}
\includegraphics[width=\columnwidth]{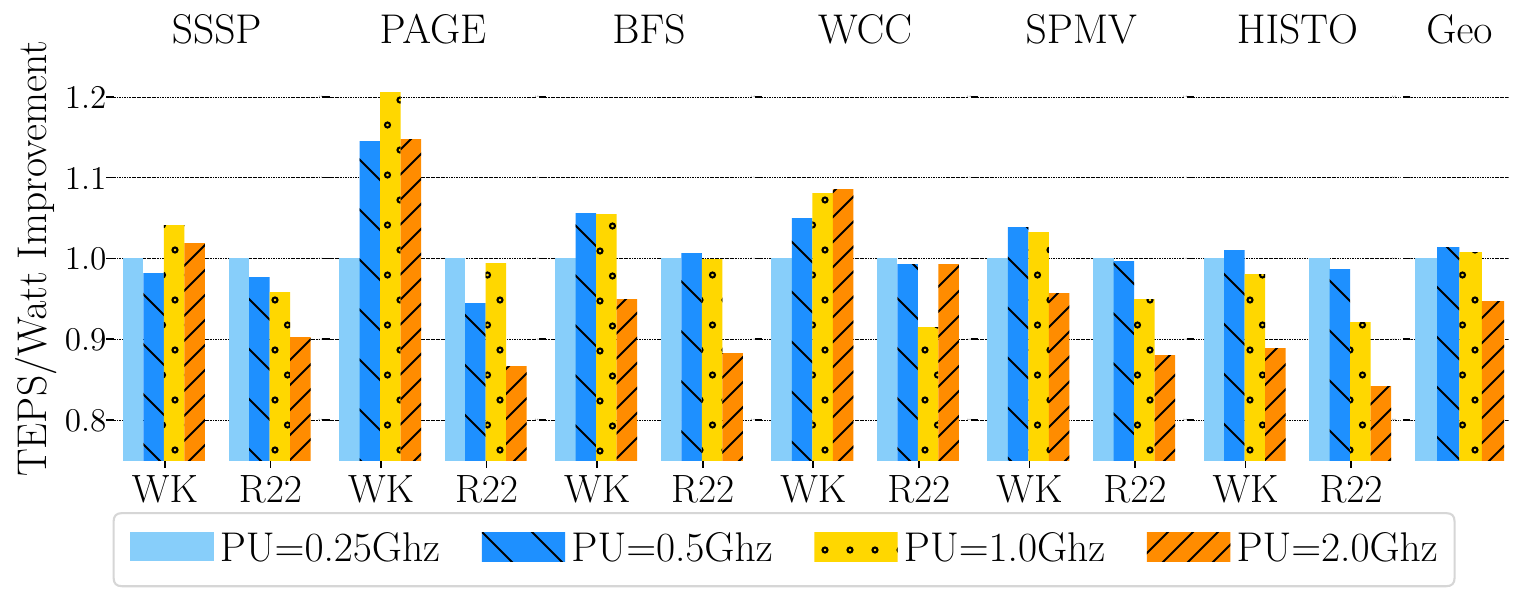}
\vspace{-5mm}
\caption{
Performance and energy efficiency improvements for different PU frequencies, over the baseline of 250 MHz.
All configurations use 64x64 tiles, one PU per tile, and 512KB SRAM per tile.
}
\vspace{-1mm}
\label{fig:PU_freq}
\end{figure}

\cref{fig:PU_freq} shows the performance improvement and energy-efficiency of \proj with different PU frequencies for our sparse application domain.
We already know that the dense workloads will scale well with PU frequency, as they have higher arithmetic intensity, but with this experiment, we want to see how these sparse applications behave.

\cref{fig:PU_freq} shows that the performance improvement is linear with PU frequency until the 1GHz point, where it starts to saturate for most applications: 2 Ghz achieves a geomean 38\% improvement over 1 Ghz.
It is noteworthy that WCC achieves full linear scaling up to 2 Ghz for R22, showing that it is not constrained by memory or the network on this scaling level (64x64 PUs).

\textbf{\textit{Default and sparse-dense:}}
The default maximum PU frequency of \proj is 1 GHz. However, if the target deployment considers the possibility of processing compute-bound applications, the 2 Ghz should be chosen, because the area overhead (assuming a pessimistic 50\% area increase) in the overall \proj die is 10\%.
The decrease in energy efficiency with the 2 Ghz version can be mitigate at runtime by using DVFS to reduce the frequency and thus power.

\subsection{Target-metric-dependent Packaging-Time Configurations}\label{sec:res_packaging_time}

Optimization of production systems for different target metrics such as fastest time-to-solution, energy or cost is an important factor both at pre-silicon as well as at packaging.
At this moment, we have covered all of our pre-silicon options, and so we now focus on post-silicon configurations, particularly, on the tradeoffs of integrating HBM or not.

\begin{figure}[t]
\includegraphics[width=\columnwidth]{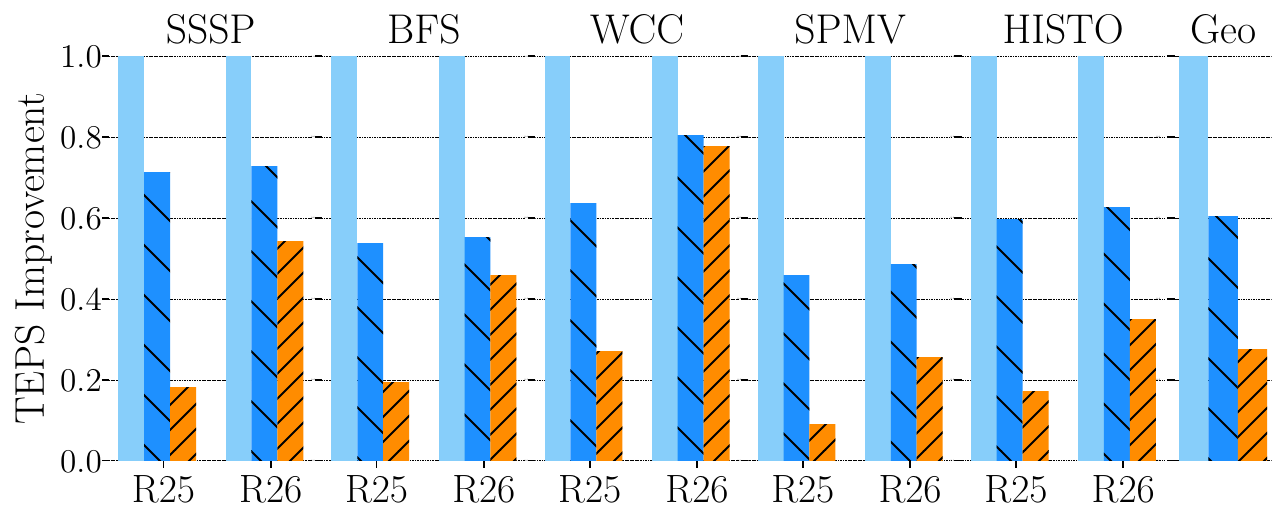}
\includegraphics[width=\columnwidth]{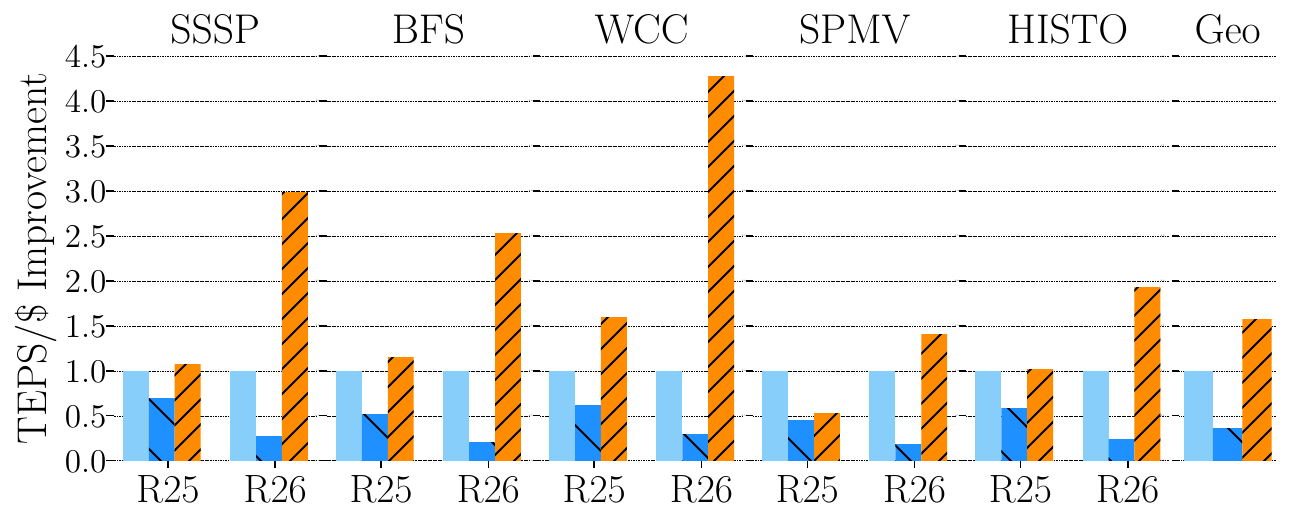}
\includegraphics[width=\columnwidth]{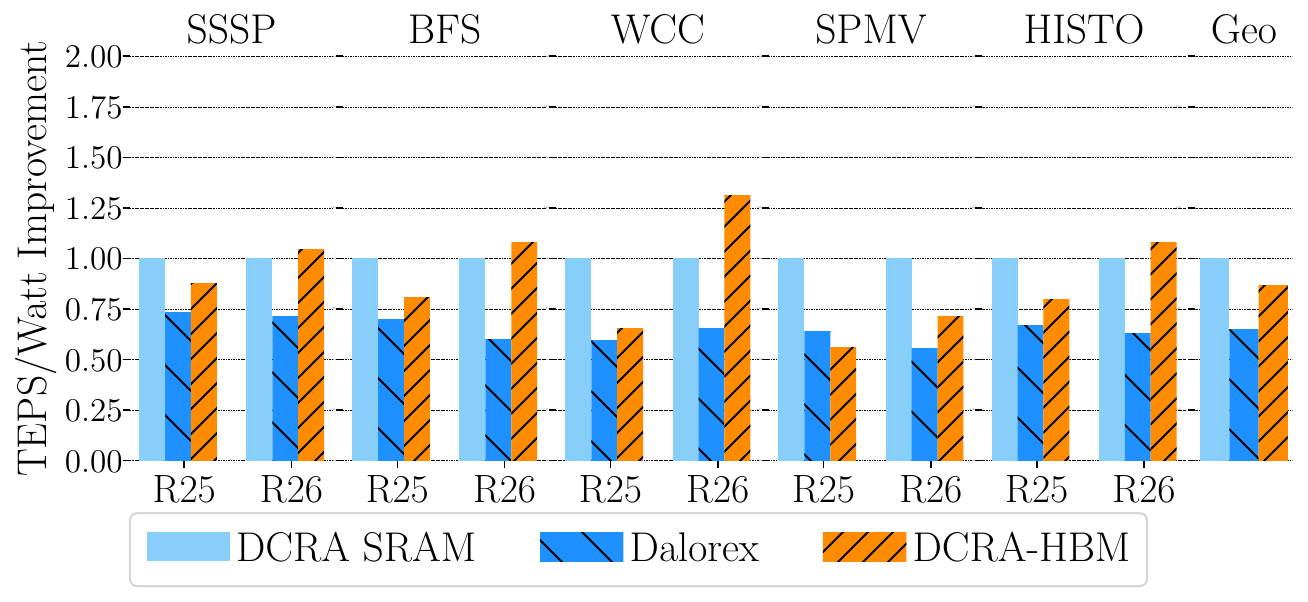}
\vspace{-6mm}
\caption{
Comparison of performance, cost-effectiveness and energy efficiency of Dalorex and \proj-HBM, over the baseline of using only \proj dies (\proj-SRAM).
Dalorex uses 2MB SRAM per tile~\cite{dalorex}, while \proj uses 512KB; \proj-HBM integrates a 8GB HBM2 device per chiplet of 32x32 tiles, providing a memory-per-PU ratio of 8MB/PU.
We parallelize the datasets across the smallest configuration for which it fits on chip.
For R25 that size is 32x32 for \proj-HBM, 64x64 for Dalorex and 128x128 for \proj-SRAM, while for R26 we use our next scaling step with 4 times as many tiles.
}
\vspace{-3mm}
\label{fig:packages_comparison}
\end{figure}

\cref{fig:packages_comparison} (above) shows the throughput of \proj with HBM, and Dalorex, over the baseline of only using \proj dies' SRAM.
Given that DCRA-SRAM, Dalorex, and DCRA-HBM have different memory capacities per tile, 512KB, 2MB and 8MB, respectively, we parallelize the datasets across the smallest configuration for which it fits on chip: 32x32 and 64x64 for \proj-HBM when running R25 and R26, respectively,  4$\times$ as many tiles for Dalorex, and 4$\times$ more for \proj-SRAM.
(For Dalorex's monolithic integration, only one chip fits in a wafer, and so we assume that for its cost.)
Given these different parallelization sizes, the most scaled-out configuration (the \proj-SRAM baseline) has the highest throughput across the board, but with varying degrees of speedups across workloads and datasets.
The gap between Dalorex and DCRA also showcases the benefits of our hierarchical Torus topology, which reduces NoC contention and latency.

In terms of throughput-per-dollar (\cref{fig:packages_comparison}, middle), the configuration including HBM wins across the board, with a couple of exceptions where the performance gains of the scaled-out \proj-SRAM configuration are enough to overcome the extra cost.
Note that the throughput-per-dollar results could change significantly if either HBM becomes cheaper (currently modeled as 7.5\$ per GB, roughly twice the cost of our selected 32x32-tile \device{} die with 512KB/die) or another DRAM or 3D-stacked SRAM~\cite{3d_sram} technology becomes cheaper.

\cref{fig:packages_comparison} (bottom) shows that \proj-SRAM wins in energy efficiency for R25 but not for R26.
We found that the energy efficiency stays more or less stable across the parallelization levels (with some ups and downs).
This trend is more clearly shown in \cref{fig:scaling_million}, and the reasons are quite fascinating:
(1) the energy of SRAM remains fairly constant across the parallelization levels, as SRAM banks are powered off when idle, as we do not need to stride arrays across them for a single PU/tile;
(2) the HBM does it similarly, to the point where it is switched off when no longer used, and the SRAM becomes the main level of memory. The memory energy becomes less dominant over time as the data cache hit-rate increases;
(3) the NoC energy increases at a similar rate at which the memory energy decreases.
(4) the PU energy remains stable since in this task-based execution model, PUs do not run a main thread but rather react to incoming tasks provided by the router (which triggers the clock gate of the PU.)

\begin{figure}[t]
\includegraphics[width=\columnwidth]{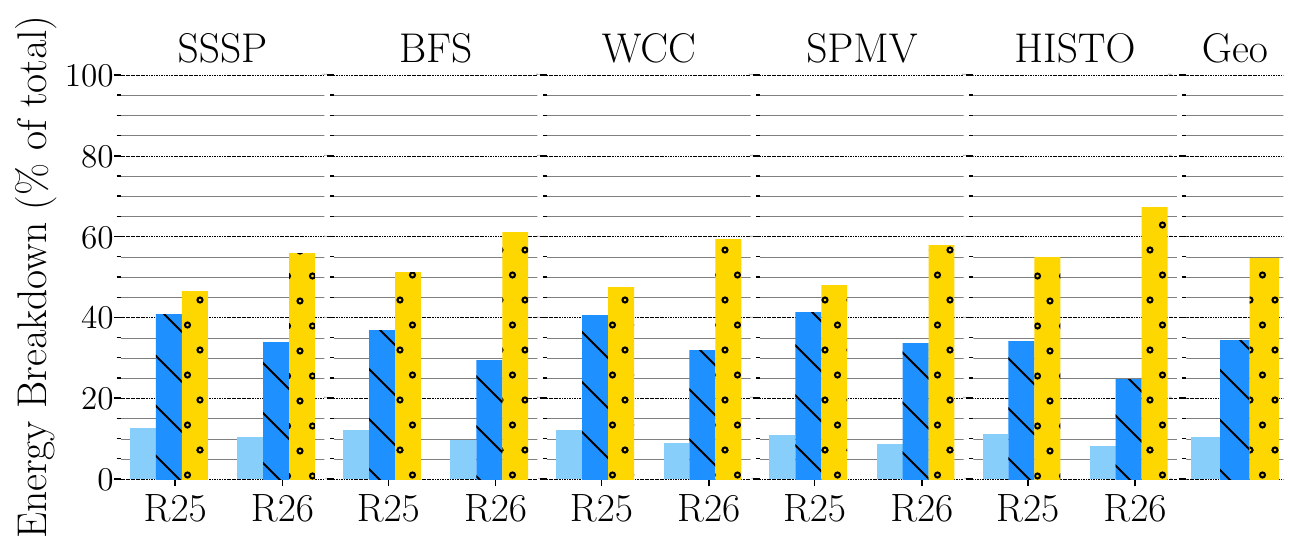}
\includegraphics[width=\columnwidth]{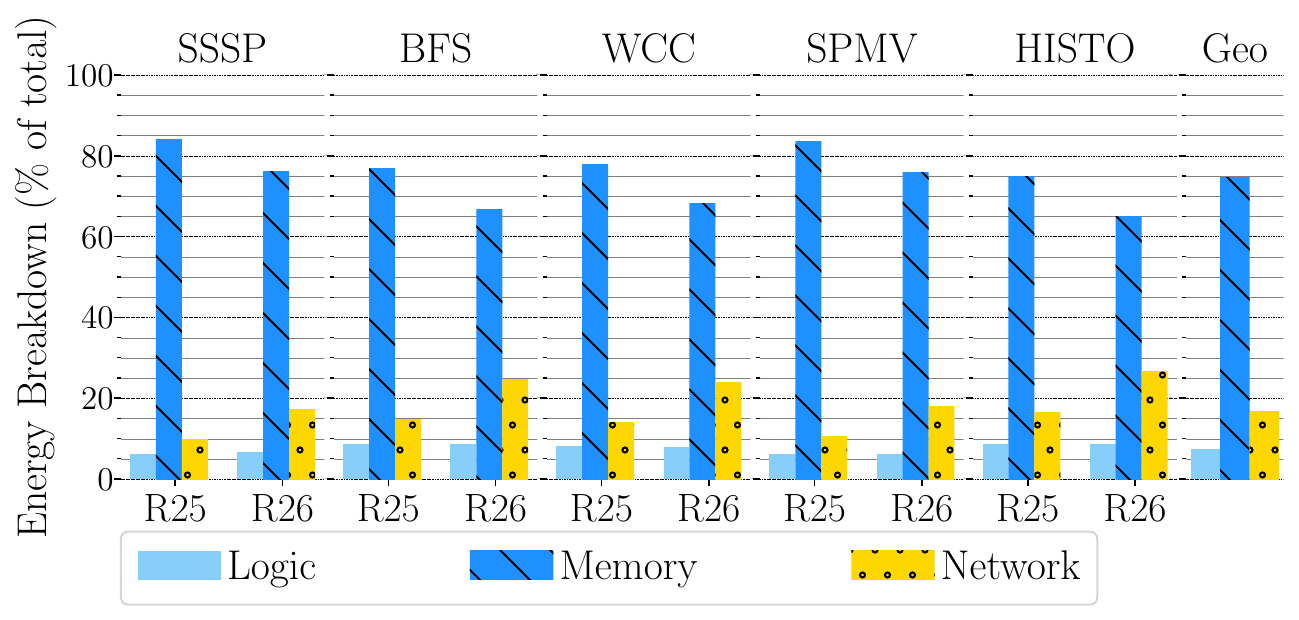}
\vspace{-6mm}
\caption{
Breakdown of the energy consumed by computing logic, memory, and NoC communication (including routing and wire energy), for Fig.\ref{fig:packages_comparison}'s \proj-SRAM (top) and \proj-HBM (bottom).
The Y-axis shows the \% of the total energy spent on each component.
}
\vspace{-2mm}
\label{fig:energy_breakdown}
\end{figure}

\cref{fig:energy_breakdown} shows the breakdown of the energy used by PUs, memory, and NoC for \proj-SRAM and \proj-HBM the configurations shown in \cref{fig:packages_comparison}.
PUs use a small fraction in both cases, partly because PUs are powered off when idle.

\textbf{\textit{\proj-SRAM energy:}}
Since this version uses 16$\times$ more tiles than \proj-HBM, it spends more energy on wires and routers (\cref{fig:scaling_million} shows the increase in message routing hops).

\textbf{\textit{\proj-HBM energy:}}
As we saw in \cref{fig:sram_configurations}, the HBM integration saturates the memory controller bandwidth, which makes the energy usage dominated by DRAM for small parallelization levels.

\subsection{Compile-time Configurations}\label{sec:million}

Once a \proj system is taped-out and packaged according to the considerations we described above, there remain several compile-time configurable features.
These include (a) the input and output queue sizes, (b) the portion of the tile's SRAM that is used as cache and a scratchpad, (c) the PU and NoC frequency (up to the maximum frequency considered during design), and (d) the network topology.

\ifarxiv
SRAM and frequency were discussed in the sections above; we would like to add that the part of the SRAM that is not used for caching space is invested for program memory, data arrays (mapped as a scratchpad), or larger queues.
This last option is the one we focus on in this section, which can have a different impact across applications and datasets.

\textit{\textbf{Queue sizes}} are especially important in the task-based execution model that we use because the parallelism achieved is equal to the number of PUs with tasks ready to execute in the IQs.
To provide enough work for others, the OQs should be sufficient.
\cref{fig:queue_characterization} shows the performance gained from increasing the size of the OQ of the task that pushes the vertex updates for the graph applications (or the updates to the output array in SPMV and Histogram) in the Dalorex implementation that we use.
We increase the size of this OQ2 over the size of the OQ for the first task (OQ1), which pushes the lookup for the edge list.
Because the ratio of the number of task invocations that go to each of these tasks is similar to the average number of edges per vertex, we can use it to choose the size of OQ2.
\cref{fig:queue_characterization} showcases that the performance improvement is more significant for R22 than for WK (which only improves for SPMV).
R22 has 32 edges per vertex on average, while WK has 25.
We observe that this ratio is also very different across applications, so this is another learning we can use to guide this decision.
A more thorough study varying the size of OQ1, as well as that of the IQs for a particular application, would help optimize this software configuration parameter. 

\begin{figure}[t]
\includegraphics[width=\columnwidth]{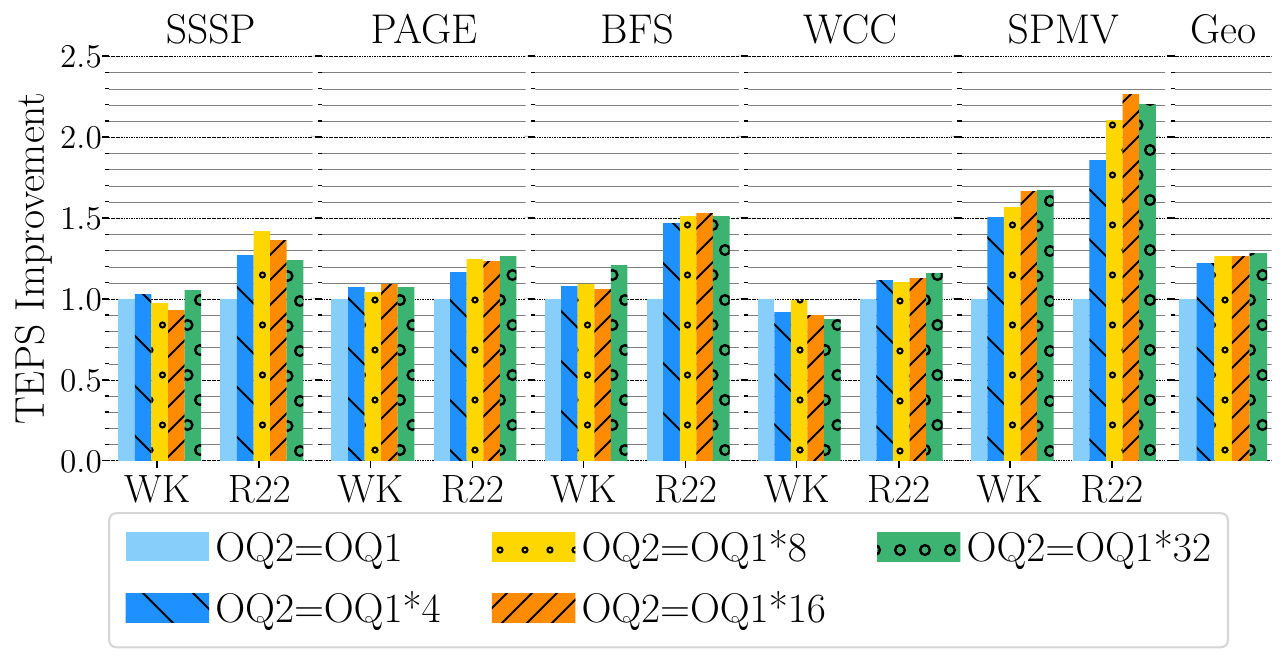}
\vspace{-5mm}
\caption{Performance improvement of increasing the size of the OQ2 over that of OQ1, normalized over the baseline of OQ2=OQ1.
OQ1 size is 12 task invocation messages.
All configurations use 64x64 tiles, one PU/tile, and 512KB/tile.
We do not show Histogram because it only has two tasks (one OQ between them).
}
\vspace{-5mm}
\label{fig:queue_characterization}
\end{figure}

\fi

\begin{figure}[t]
\hspace{-3mm} \includegraphics[width=1.05\columnwidth]{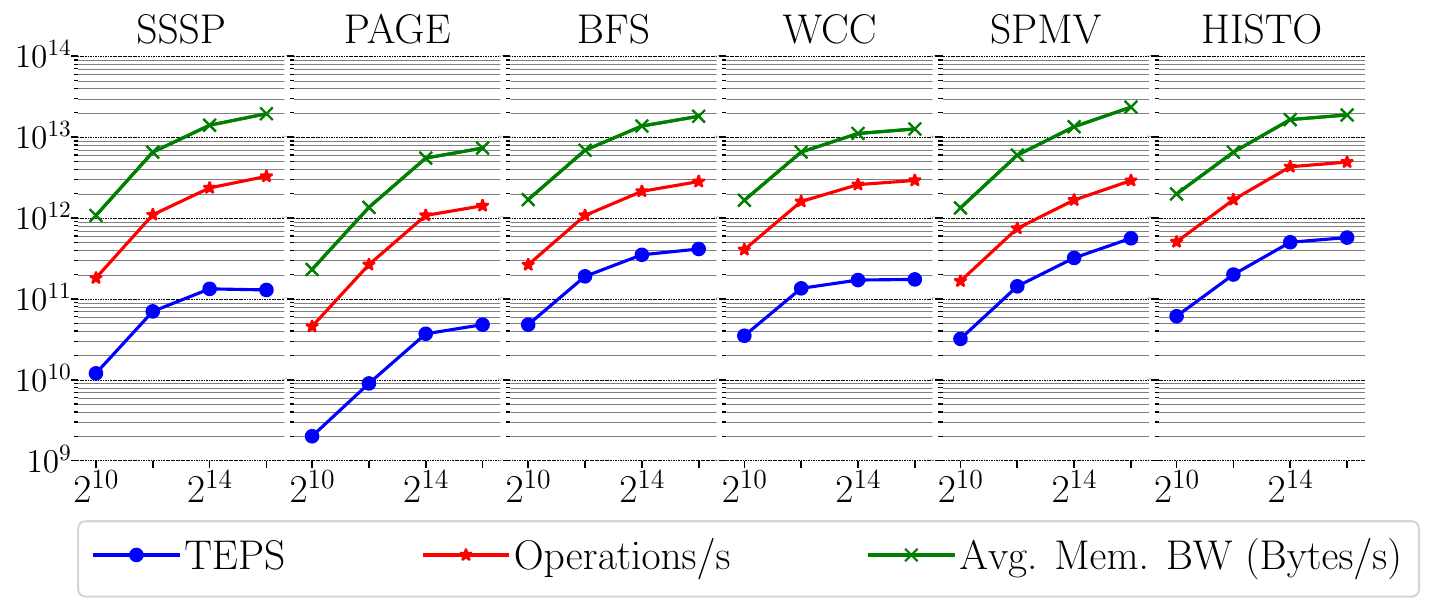}
\vspace{+1mm}
\hspace{-3mm} \includegraphics[width=1.02\columnwidth]{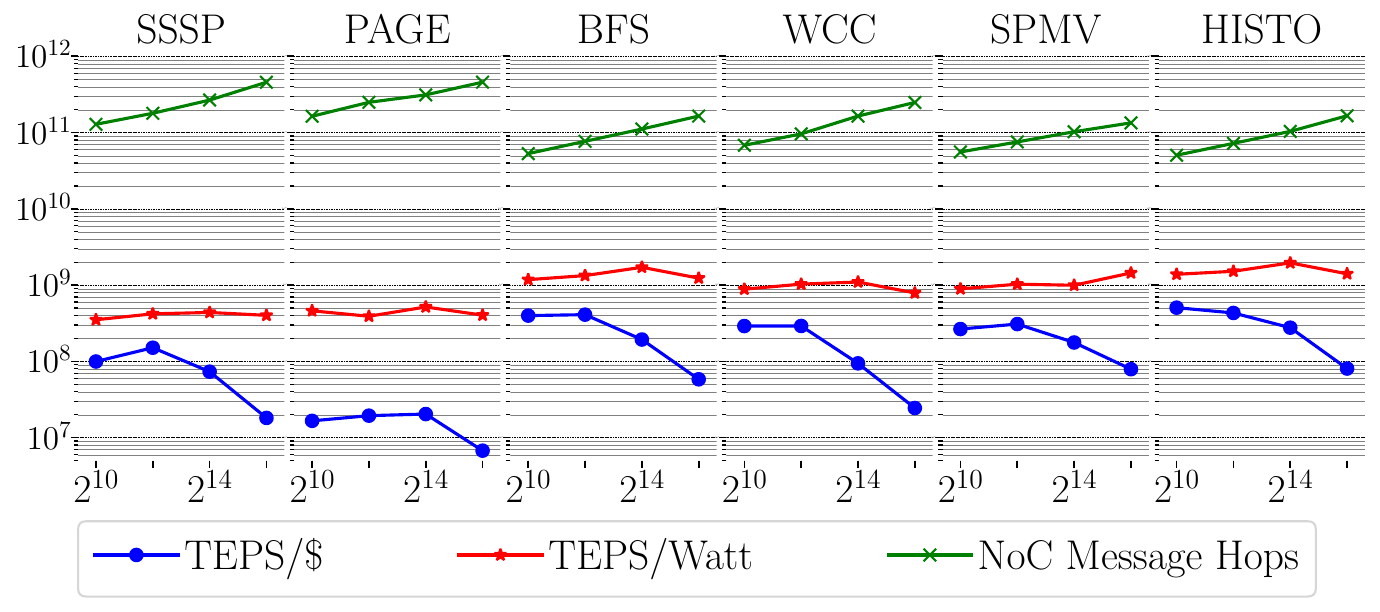}
\vspace{-7mm}
\caption{
Throughput in operations and floating point operations (FLOPs) per second, and the average on-chip memory bandwidth needed to achieve that.
The X-axis is the size of the \proj grid used when analyzing strong scaling R26, ranging from 1024 to over 64K PUs.
The bottom plot shows throughput as a function of power and cost, in addition to the total number of message hops during the execution.
}
\vspace{-4mm}
\label{fig:scaling_million}
\end{figure}

\textit{\textbf{Parallelization Level:}}
An important compile-time decision is, of course, the size of the subgrid of tiles that we use to parallelize the application, from the total number of tiles that the system has. 
This is related to the NoC topology because the configurable routers described in \cref{fig:torus} need to be set to wrap around accordingly.
The 2D Torus can be as large as a computing node (a board of one or more packages).
Beyond that level, the nodes are connected via the cluster-level network.

Evaluating how the performance scales with the size of the node can help us understand the tradeoffs of using a larger or smaller node.
\cref{fig:scaling_million} (top) shows the throughput and the average memory bandwidth needed to achieve that, for R26.
The gap between the floating-point and regular operations shows the characteristics of the application, as well as the gap between the memory bandwidth and the throughput.
Analyzing \cref{fig:scaling_million}, we learn the following principles that guide our decision tree of \cref{fig:flowchart}.

\textbf{\textit{Strong scaling for faster time-to-solution:}}
We parallelize a dataset with $2^{26}$ vertices with up to $2^{16}$ tiles.
At that level of parallelization, the entire dataset fits on the 32GB of SRAM across all 16 chips.
\cref{fig:scaling_million} (bottom, green line) shows that the total number of message hops increases as we parallelize across more tiles.
This means that to do the same amount of total work, the PUs need to communicate more.
Although the throughput increases until this point, we do not expect it to scale much further without applying communication-mitigation techniques like graph partitioning or reduction trees.

\textbf{\textit{Throughput-per-watt remains stable until 64K tiles:}}
\cref{fig:scaling_million} (bottom, red line) shows that the throughput-per-watt remains stable across different parallelization levels.
From \cref{fig:energy_breakdown} we observe that as we scale out, the energy consumption shifts from HBM towards the NoC as the number of tiles increases.
This is because the HBM device is powered down when the entire dataset fits in SRAM (at the last scaling point); SRAM then behaves as the main memory level of the chip.
Beyond this point, we can only expect the efficiency to drop, as the traffic continues to increase with the number of tiles.

\textbf{\textit{Throughput-per-dollar likes small grids:}}
\cref{fig:scaling_million} (bottom, blue line) shows that if throughput-per-dollar is the target metric, staying within $2^{12}$ tiles (64x64) is the best option.
This is because we stop seeing linear speedups beyond $2^{12}$, while the cost always grows linearly.
Note that while the TEPS/\$ trend is determined by the sublinear scaling, the absolute values depend on the cost.
While we can power down HBM to save energy, we cannot avoid the HBM cost when we do not use it.
Thus, the $2^{16}$ datapoint, for which the dataset fits on SRAM, would achieve a higher TEPS/\$ than the one displayed if we had chosen an integration without HBM.
However, as we saw in \cref{fig:packages_comparison}, TEPS/\$ for SRAM-only at $2^{16}$ is still lower than staying within the $2^{12}$ scaling point with HBM.

\section{Discussion: Choosing Among Options and Configurations}\label{sec:discussion}

\begin{figure}
\centering
\includegraphics[width=\columnwidth]{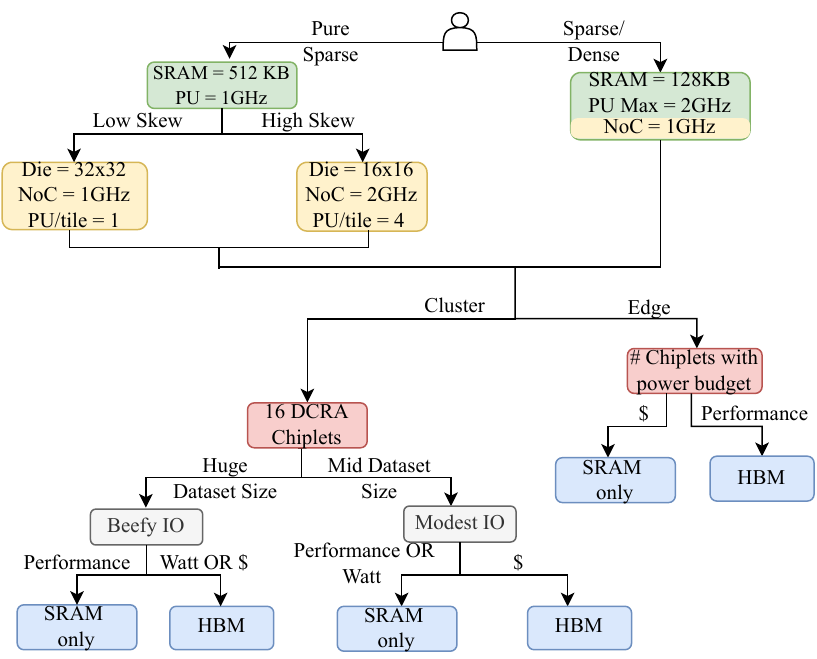}
\vspace{-5mm}
\caption{
Decision diagram to choose among the options and configurations of \proj based on the target deployment and metrics of interest.
}
\vspace{-4mm}
\label{fig:flowchart}
\end{figure}

\cref{fig:flowchart} exemplifies what the decision process to design a \proj-based system could look like, with five variables:
(1) target application domain, e.g., pure-sparse or a balance between sparse and dense;
(2) skewness of the data;
(3) target deployment, e.g., HPC cluster or edge device;
(4) dataset scale;
(5) target metrics, e.g., fastest time-to-solution, energy, or cost.

Based on the target application domain, we decide whether to have a max operating frequency of 1Ghz or 2Ghz, and whether to employ 512KB of SRAM or 128KB, respectively.
In addition, a high-skew target datasets (that create execution hotspots), choosing four PUs per tile consuming from the same task queue can help mitigate that (in that case, we double the NoC frequency to balance the injection throughput to the tiles).

Depending on whether the target is an HPC cluster or an edge deployment, HBM and SRAM can be integrated differently.
The HPC tradeoffs were discussed in \cref{sec:million}, but we expect the edge-device tradeoffs to be different.
In the HPC cluster, we could scale out to use more chips, whereas in the edge we may be more constrained by size and power.
At the edge, if we care most about performance we should integrate HBM, and the cost-efficient alternative is to use SRAM, and when the dataset does not fit we have to resort to swapping back and forth from DDR or flash.
Additionally, the IO bandwidth between chips and outside the node could be provisioned depending on whether the cluster is expecting to run huge graphs collectively, or every node processing its graph.
For that, graph partitioning techniques play an important role.

\section{Conclusion}\label{sec:conclusions}
\vspace{-1mm}

\proj proposes a chiplet-based architecture that allows making key design decisions, such as on-chip memory and off-chip interconnect, post-silicon.
This enables the same \device{} design to be mass-produced (saving NRE costs) and later integrated differently to create products with distinct target metrics.
We have discussed the configurations that warrant different targets---some of which are evaluated in \cref{sec:results}. 
To enable such configurations, \proj includes our design of a 2D-Torus NoC topology which can span across a varying number of dies as its size is defined in software.

We put much effort into designing a chiplet-based architecture that is highly configurable, allowing it to potentially be applied to graph and sparse application domains as well as others to optimize the chip at packaging time for different target metrics, such as time-to-solution, energy efficiency, or cost.
\cref{fig:flowchart} summarizes our guidelines for integrating DCRA for different use cases and constraints. 

We have created the scripts for the artifact evaluation of the experiments presented in this paper.
Our \repository includes the simulation framework implementing \proj; we are planning to document it with the publication of this paper to allow researchers to explore further configurations and optimizations of \proj.